\documentclass[acmsmall,screen,nonacm]{acmart}
\usepackage{color}
\usepackage{listings}
\usepackage{hyperref}
\usepackage{amsmath,amsfonts}
\usepackage{mathtools}
\usepackage[ruled, vlined, linesnumbered]{algorithm2e}
\SetAlFnt{\small}
\SetAlCapFnt{\small}
\SetAlCapNameFnt{\small}
\usepackage{multirow}
\usepackage{graphicx}
\usepackage{makecell}
\usepackage{caption}
\usepackage{subcaption}
\usepackage{mathpartir}
\usepackage{adjustbox}
\usepackage{mdframed}
\usepackage{siunitx}

\usepackage{pifont}
\newcommand{\gcmark}{\textcolor{green}{\textbf{\ding{51}}}}
\newcommand{\rxmark}{\textcolor{red}{\textbf{\ding{55}}}}

\usepackage{wasysym}

\usepackage[frozencache=true,cachedir=minted-cache]{minted} 
\newlength{\textfloatsepsave}
\usepackage{xcolor}

\makeatletter
\let\old@lstKV@SwitchCases\lstKV@SwitchCases
\def\lstKV@SwitchCases#1#2#3{}
\makeatother
\usepackage{lstlinebgrd}
\makeatletter
\let\lstKV@SwitchCases\old@lstKV@SwitchCases

\lst@Key{numbers}{none}{%
    \def\lst@PlaceNumber{\lst@linebgrd}%
    \lstKV@SwitchCases{#1}%
    {none:\\%
     left:\def\lst@PlaceNumber{\llap{\normalfont
                \lst@numberstyle{\thelstnumber}\kern\lst@numbersep}\lst@linebgrd}\\%
     right:\def\lst@PlaceNumber{\rlap{\normalfont
                \kern\linewidth \kern\lst@numbersep
                \lst@numberstyle{\thelstnumber}}\lst@linebgrd}%
    }{\PackageError{Listings}{Numbers #1 unknown}\@ehc}}
\makeatother

\newcommand\code[1]{\mintinline{python}{#1}}
\newcommand{\Name}{{Allo}\xspace}
\newcommand{\MLIR}{{MLIR}\xspace}
\newcommand{\LLVM}{{LLVM}\xspace}
\newcommand{\RNum}[1]{\uppercase\expandafter{\romannumeral #1\relax}}

\DeclareMathOperator*{\argmin}{arg\,min}

\newcommand{\showcomments}{yes}
\newcommand\fixme[1]{
    \ifthenelse{\equal{\showcomments}{yes}}{\textcolor{red}{\small [FIXME: #1]}}{\ignorespaces}
}
\newcommand\hz[1]{
    \ifthenelse{\equal{\showcomments}{yes}}{\textcolor{red}{\small [hz: #1~]}}{\ignorespaces}
}
\newcommand\ns[1]{
    \ifthenelse{\equal{\showcomments}{yes}}{\textcolor{purple}{\small [ns: #1~]}}{\ignorespaces}
}
\newcommand\sj[1]{
    \ifthenelse{\equal{\showcomments}{yes}}{\textcolor{green}{\small [sj: #1~]}}{\ignorespaces}
}
\newcommand\zz[1]{
    \ifthenelse{\equal{\showcomments}{yes}}{\textcolor{blue}{\small [zz: #1]}}{\ignorespaces}
}

\AtBeginDocument{%
  }

\begin{document}

\title{\Name: A Programming Model for Composable Accelerator Design}


\author{Hongzheng Chen}
\authornote{Equal contribution.}
\orcid{0000-0002-6617-0075}
\affiliation{%
  \institution{Cornell University}
  \country{USA}
}
\email{hzchen@cs.cornell.edu}

\author{Niansong Zhang}
\authornotemark[1]
\orcid{0000-0002-2850-0176}
\affiliation{%
  \institution{Cornell University}
  \country{USA}
}
\email{nz264@cornell.edu}

\author{Shaojie Xiang}
\orcid{0000-0002-6901-8837}
\affiliation{%
  \institution{Cornell University}
  \country{USA}
}
\email{sx233@cornell.edu}

\author{Zhichen Zeng}
\orcid{0009-0005-0023-2367}
\authornote{Work was done when Zhichen and Mengjia interned at Cornell.}
\affiliation{%
  \institution{University of Science and Technology of China}
  \country{China}
}
\email{zhichenzeng@mail.ustc.edu.cn}

\author{Mengjia Dai}
\orcid{0000-0002-7347-5117}
\authornotemark[2]
\affiliation{%
  \institution{University of Science and Technology of China}
  \country{China}
}
\email{mjd20021014@mail.ustc.edu.cn}

\author{Zhiru Zhang}
\orcid{0000-0002-0778-0308}
\affiliation{%
  \institution{Cornell University}
  \country{USA}
}
\email{zhiruz@cornell.edu}


\begin{abstract}
Special-purpose hardware accelerators are increasingly pivotal for sustaining performance improvements in emerging applications, especially as the benefits of technology scaling continue to diminish. 
However, designers currently lack effective tools and methodologies to construct complex, high-performance accelerator architectures in a productive manner. Existing high-level synthesis (HLS) tools often require intrusive source-level changes to attain satisfactory quality of results. Despite the introduction of several new accelerator design languages (ADLs) aiming to enhance or replace HLS, their advantages are more evident in relatively simple applications with a single kernel. Existing ADLs prove less effective for realistic hierarchical designs with multiple kernels, even if the design hierarchy is flattened.

In this paper, we introduce \Name, a composable programming model for efficient spatial accelerator design.
\Name decouples hardware customizations, including compute, memory, communication, and data type from algorithm specification, and encapsulates them as a set of customization primitives.
\Name preserves the hierarchical structure of an input program by combining customizations from different functions in a bottom-up, type-safe manner. This approach facilitates holistic optimizations that span across function boundaries.
We conduct comprehensive experiments on commonly-used HLS benchmarks and several realistic deep learning models.
Our evaluation shows that \Name can outperform state-of-the-art HLS tools and ADLs on all test cases in the PolyBench.
For the GPT2 model, the inference latency of the \Name generated accelerator is 1.7$\times$ faster than the NVIDIA A100 GPU with 5.4$\times$ higher energy efficiency, demonstrating the capability of \Name to handle large-scale designs.
\end{abstract}




\maketitle

\section{Introduction}
\label{sec:intro}
With the recent trends in technology scaling, computer engineers are increasingly turning to special-purpose hardware accelerators to meet the escalating computational demands of emerging applications, such as large language models (LLMs)~\cite{radford2019gpt2,openai2023gpt4,touvron2023llama}.
One architectural paradigm that has gained popularity is spatial architecture~\cite{jouppi2017tpuv1,jouppi2023tpuv4,emani2021sambanova,genc2021gemmini,markidis2018tensorcore,xilinxAIE}, which instantiates specialized processing engines interconnected through direct wires or streaming buffers to increase throughput and reduce off-chip memory accesses. 
While hardware specialization can significantly improve performance and energy efficiency, it does entail a substantially higher development effort. Specifically, manually constructing spatial architectures has been notably challenging, particularly with the traditional register-transfer-level (RTL) design abstraction. Consequently, modern accelerator designs are increasingly embracing high-level synthesis (HLS) to expedite RTL code generation and enable rapid exploration of diverse design alternatives~\cite{cong2011fpgahls,cong2022fpgahls,lai2021fpgasurvey}.
However, to achieve high performance, HLS users must extensively restructure the source program to guide the tool toward realizing specialized architectures like systolic arrays. Additionally, they are required to employ various vendor-specific data types and pragmas, diminishing design reusability and portability.

In this context, we identify two major challenges to the productive development of high-performance accelerators.

\textbf{Challenge 1: Balancing manual control with automated compiler optimizations.}
Kernels manually created by experts deliver high-performance implementations but require substantial manual effort for design and validation. Also, these kernels usually adhere to specific data types and function signatures, which hampers their ability to keep up with rapidly evolving applications and hardware advancements. There is an increasing use of automated compiler techniques such as polyhedral compilation to generate on-chip buffers~\cite{pouchet2013buffer}, streaming dataflow architectures~\cite{chi2018soda}, or systolic arrays~\cite{jason2018polysa,wang2021autosa} from a plain C/C++ code without sophisticated loop annotations.
However, these tools typically do not provide adequate control to the designers to explore various performance/cost trade-offs and customize the memory hierarchies and communication schemes for new applications.
Embracing domain-specific languages (DSLs) simplifies tasks for both programmers and compilers~\cite{fahim2021hls4ml,milder2012spiral,duarte2018particle,steward2018ripl,hegarty2014darkroom},
but most DSLs are inherently tailored for specific application domains, such as image processing, machine learning, and network processing, and they lack support for general-purpose language constructs essential for accelerator hardware design~\cite{lai2021fpgasurvey}.


\textbf{Challenge 2: Bridging the gap from single-kernel optimization to complex multi-kernel accelerator design.}
In recent trends, DSLs for hardware design are evolving to become more generalized, incorporating flexible imperative language constructs or being embedded in general host languages such as C++, Python, or Scala~\cite{lai2019heterocl,nitish2019t2s,huang2021pylog,koeplinger2018spatial,rachit2020dahlia}.
We refer to this category of programming models as accelerator design languages (ADLs). Inspired by Halide~\cite{jrk2013halide} and TVM~\cite{chen2018tvm}, several recent ADLs further separate algorithm definition from hardware optimizations~\cite{lai2019heterocl,nitish2019t2s}, which improves both productivity and portability. However, existing ADLs primarily focus on optimizing single application kernels like convolution and matrix multiplication. In the case of realistic multi-kernel applications, these ADLs tend to generate monolithic flattened designs, sidestepping the intricacies of composing distinct kernels, which may present incompatible interfaces or conflicting optimizations. The inadequate support for composability compromises modularity, debuggability, and often leads to suboptimal performance, as pre-optimized kernels cannot easily be integrated into a hierarchical program structure. 

To tackle these challenges, we propose \emph{\Name}, a new programming model for composable design of high-performance spatial accelerator architectures\footnote{Allo means atypical, which reflects our emphasis on creating non-traditional hardware architectures. The framework is open-source: \url{https://github.com/cornell-zhang/allo}.}.
The key design principles of \Name are to provide decoupled hardware customization primitives, modularize the accelerator design process, and facilitate type-safe composition of individual components.


\noindent\textbf{Progressive Hardware Customizations.}
We inherit the idea from popular schedule languages like TVM~\cite{chen2018tvm} and Halide~\cite{jrk2013halide} to decouple hardware customizations (e.g., caching and pipelining) from algorithm specifications.
Each hardware customization is a primitive that performs a rewrite on the program.
We not only decouple the loop-based transformations, but also extend the decoupling to memory, communication, and data types.
Each customization primitive can be verified individually and progressively applied to a vanilla program to conduct optimizations.

\noindent\textbf{Reusable Parameterized Kernel Templates.}
\Name supports declaring type variables during kernel creation and instantiating the kernel when building the hardware executable, which is a feature absent in most hardware ADLs~\cite{lai2019heterocl,huang2021pylog,koeplinger2018spatial} but is important for building reusable hardware kernel libraries.
\Name introduces a concise grammar for creating kernel templates, eliminating the need for users to possess complicated metaprogramming expertise.


\noindent\textbf{Composable Schedules.}
\Name empowers users to construct kernels incrementally from the bottom up, adding customizations one at a time while validating the correctness of each submodule.
Ultimately, multiple schedules are progressively integrated into a complete design using the \code{.compose()} primitive.
This approach, unachievable by prior top-down methods, significantly enhances productivity and debuggability.

\noindent\textbf{Holistic Dataflow Optimizations.}
We introduce a hierarchical dataflow graph to support the composition of multiple kernels within a complex design while maintaining the function boundaries. To ensure the correctness of the interfaces when integrating distinct kernels, we model the interface unification problem as a type inference problem and solve it efficiently through dataflow analysis.
Leveraging the hierarchical dataflow graph, we can effectively size the streaming buffers (FIFOs) between stages.

To improve the usability of \Name, we have implemented the frontend language in Python, allowing for a flexible programming style with minimal type annotations.
We also present an end-to-end optimizing compiler for \Name, allowing users to write Python programs and generate the hardware bitstream.
Moreover, we provide an \MLIR dialect that supports decoupled hardware customizations at the IR level and potentially supports multiple different input languages.
In summary, our contributions are as follows:
\begin{itemize}
\item We introduce \Name, a composable programming model that enables progressive hardware customizations, transforming a vanilla program into a high-performance design, with each step being verifiable.
\item We propose composable schedules, enabling users to construct modular hardware accelerators from the ground up by combining customized kernels and external IPs.
A type system for the memory layout is also proposed to ensure type safety during schedule composition.
Additionally, we introduce holistic dataflow optimizations to ensure functional correctness and enhance performance further.
\item We conduct comprehensive experiments on both realistic benchmarks and large neural networks.
For PolyBench~\cite{pouchet2012polybench}, we outperform several state-of-the-art HLS tools and ADLs~\cite{yehpca2022scalehls,lai2019heterocl,huang2021pylog,rachit2020dahlia}, across all design cases.
Furthermore, we demonstrate the applicability of our programming model in the context of large neural network designs.
To the best of our knowledge, we are the first to employ such an ADL for a complete evaluation of LLMs on an FPGA.
Our experimental results reveal a 1.7$\times$ speedup and 5.4$\times$ higher energy efficiency on the GPT2 model compared to the A100 GPU.
\end{itemize}

\section{An \Name Example}
\begin{figure}[t]
\begin{subfigure}[b]{0.33\linewidth}
\begin{minted}[linenos,
               fontsize=\scriptsize,
               xleftmargin=1.8em,
               autogobble]{python}
import allo
from allo.ir.types import float32
# Algorithm specification
M, N, K = 32, 32, 32
def gemm(A: float32[M, K],
         B: float32[K, N],
         C: float32[M, N]):
 for i in range(M):
  for j in range(N):
   for k in range(K):
    C[i, j] += A[i, k] * B[k, j]

# Schedule construction
s = allo.customize(gemm)
s.unroll("k", 8)
s.partition(A, dim=1, factor=8)
s.partition(B, dim=0, factor=8)

# Codegen
s.build(target="hls")
\end{minted}
\caption{An example \Name program}
\label{subfig:allo}
\end{subfigure}
\hfill\vline\hfill
\begin{subfigure}[b]{0.33\linewidth}
\begin{minted}[fontsize=\scriptsize,
               escapeinside=||]{cpp}
module {
 func.func |\textcolor{purple}{@gemm}|(
    %arg0: memref<32x|\textcolor{orange}{32}|xf32>,
    %arg1: memref<32x|\textcolor{orange}{32}|xf32>,
    %arg2: memref<32x|\textcolor{orange}{32}|xf32>) {
 // Algorithm specification
 affine.for %arg2 = 0 to 32 {
  affine.for %arg3 = 0 to 32 {
   affine.for %arg4 = 0 to 32 {
    // ... Computation
   } {loop_name = "k"}
  } {loop_name = "j"}
 } {loop_name = "i", op_name = "Sijk"}
 %lk = allo.loop_handle "Sijk", "k"
 // Decoupled customizations
 allo.unroll(%lk)
 allo.partition(%arg0, 1, 8)
 allo.partition(%arg1, 0, 8)
 return
}}
\end{minted}
\caption{The corresponding \MLIR code}
\label{subfig:mlir}
\end{subfigure}
\hfill\vline\hfill
\begin{subfigure}[b]{0.3\linewidth}
\begin{minted}[fontsize=\scriptsize]{cpp}
void gemm(
 float A[32][32], float B[32][32],
 float C[32][32]) {
 #pragma partition var=A cyclic \
   factor=8 dim=1
 #pragma partition var=B cyclic \
   factor=8 dim=0
 for (int i = 0; i < 32; ++i) {
   for (int j = 0; j < 32; ++j) {
    for (int k = 0; k < 32; ++k) {
    #pragma unroll factor=8
    // ...
}}}}}
\end{minted}
\centering
\includegraphics[width=0.85\linewidth]{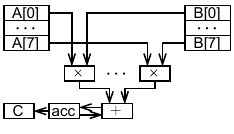}
\caption{The generated HLS code}
\label{subfig:hls}
\end{subfigure}
\caption{\textbf{An example \Name program and the corresponding \MLIR and C++ code} --- The code snippets are simplified for demonstration purposes.}
\label{fig:allo_gemm}
\Description{\textbf{An example \Name program and the corresponding \MLIR and C++ code} --- The code snippets are simplified for demonstration purposes.}
\end{figure}
Existing HLS tools often demand users to restructure their application code and insert vendor-specific data types and pragmas to achieve high performance, which are not portable and maintainable.
Moreover, with the prevalent use of Python-based frameworks~\cite{paszke2019pytorch,wolf2019huggingface} for deep learning models, manually translating those models into HLS C++ is impractical.
Therefore, we emphasize the following features as key principles when designing \Name:
(1) \textbf{Pythonic}: embracing the Python ecosystem makes the \Name coding experience similar to using native Python and effectively reduces the learning burden;
(2) \textbf{Separation of concerns}: decoupled hardware customizations make the high-performance programs easier to write and maintain;
and (3) \textbf{Composability}: all the kernels, primitives, and schedules should be composable to form complex designs.

In the following, we begin by implementing a general matrix multiplication (GEMM) kernel in \Name to illustrate the basic syntax and provide clues on why \Name can offer greater productivity compared to HLS C++.
As shown in Fig.~\ref{subfig:allo}, we first define the algorithm specification of the GEMM kernel (Lines 5-11), which specifies \emph{what} the kernel computes.
Since \Name is a Python-embedded programming language, it supports all the imperative grammars in Python (e.g., \code{if}-\code{else}, \code{for}, and \code{while}), with the distinction that users must provide explicit type annotations for function arguments and variable declaration.
This requirement arises from the dynamic typing nature of Python, which may not be inherently suitable for hardware generation where static data types are necessary to determine the accurate data bitwidth.
The type annotation in \Name consists of the basic element types and shapes.
Formal definitions of the types can be found in Supplementary Material \textcolor{ACMPurple}{A}.
Apart from the native integer and floating-point data types in Python, \Name accommodates arbitrary-bitwidth integer and fixed-point types.
This generality is important for designing high-performance accelerators that declare bitwidth only as needed, ensuring adaptability to diverse hardware requirements.

Once the algorithm is specified, we create a \emph{schedule} by calling \code{allo.customize} (Line 14).
The function passed into \code{.customize()} is treated as an \Name kernel and will be parsed by the \Name compiler.
The schedule is a sequence of optimizations, which specifies \emph{how} the kernel is executed on real hardware.
These optimizations can be applied to different algorithms and are independent of any specific algorithm, which allows us to decouple them from the algorithm and encapsulate each customization as a primitive.
We unroll the innermost loop by a factor of 8 and provide multiple banks for array $A$ and $B$ for parallel access using the provided primitives (Lines 15-17).
\Name utilizes \MLIR as the intermediate representation (IR) and provides an \MLIR dialect to decouple these hardware customizations at the IR level, as shown in Fig.~\ref{subfig:mlir}.

Lastly, we call \code{s.build} (Line 20) to lower the \MLIR module to the target backend, generating the HLS code as depicted in Fig.~\ref{subfig:hls}.
The inserted pragmas align with the schedule in the frontend program, and the generated accelerator executes the GEMM kernel with a parallelism factor of 8.

\section{Pitfalls in HLS-Based Hardware Accelerator Design}


In this section, we delve deeper into the limitations of existing HLS tools, which motivates the design of \Name.
We identify two common pitfalls in HLS and conduct several experiments to demonstrate these issues.
For the experiments, we use a widely used commercial HLS tool and target the AMD Alveo U280 FPGA~\cite{u280} with a frequency set to \SI{300}{MHz}.

\subsection{Single-Kernel Design}
\label{sub:motivation_single}
\begin{figure}[t]
\begin{minipage}{0.55\linewidth}
\begin{minted}[linenos,
               xleftmargin=1.8em,
               fontsize=\scriptsize,
               tabsize=2,
               escapeinside=||]{cpp}
// Vanilla
void gemm(
  float A[1024][1024], float B[1024][1024],
  float C[1024][1024]
) {
  for (int i = 0; i < 1024; ++i) {
    for (int j = 0; j < 1024; ++j) {
      for (int k = 0; k < 1024; ++k) {
        C[i][j] += A[i][k] * B[k][j];
}}}}

// Inner-product
void ip_gemm(
  float A[1024][1024], float B[1024][1024],
  float C[1024][1024]
) {
  #pragma partition var=A cyclic factor=32 dim=1
  #pragma partition var=B cyclic factor=32 dim=0
  for (int i = 0; i < 1024; ++i) {
    for (int j = 0; j < 1024; ++j) {
      for (int k = 0; k < 1024; ++k) {
      #pragma HLS pipeline |\colorbox{yellow}{II=1} \textcolor{red}{Is it achievable?}|
      #pragma HLS unroll factor=32
        C[i][j] += A[i][k] * B[k][j];
}}}}
\end{minted}
\bigskip
\resizebox{0.9\linewidth}{!}{
\scriptsize
\begin{tabular}{ccccc}\Xhline{1pt}
               & \thead{Latency (ms)} & \thead{II}  & \thead{Freq. (MHz)} & \thead{Speedup} \\\Xhline{1pt}
Vanilla        &       25074 &  7  & 427  & 1$\times$         \\
Inner-product  &       17950 & \textcolor{red}{128} & 240  & 1.4$\times$         \\
Row-wise product  &    \textbf{112} &  1  & 427 & 223$\times$   \\\Xhline{1pt}
\end{tabular}
}
\end{minipage}
\begin{minipage}{0.43\linewidth}
\begin{minted}[linenos,
               fontsize=\scriptsize]{cpp}
// Row-wise product
void rp_gemm(
  float A[1024][1024], float B[1024][1024],
  float C[1024][1024]
) {
  #pragma partition var=B cyclic 32 dim=1
  #pragma partition var=C cyclic 32 dim=1
  float buf_C[1024];
  #pragma partition var=buf_C cyclic 32 dim=0
  l_i: for (int i = 0; i < 1024; i++) {
    // 1) initialization
    l_j_init: for (int j = 0; j < 1024; j++) {
    #pragma pipeline II=1
    #pragma unroll factor=32
      buf_C[j] = C[i][j];
    }
    // 2) computation
    l_k: for (int k = 0; k < 1024; k++) {
    // reordered reduction loop
      float a = A[i][k];
      l_j: for (int j = 0; j < 1024; j++) {
      #pragma pipeline II=1
      #pragma unroll factor=32
        buf_C[j] += a * B[k][j];
    }}
    // 3) write-back
    l_j_back: for (int j = 0; j < 1024; j++) {
    #pragma pipeline II=1
    #pragma unroll factor=32
      C[i][j] = buf_C[j];
}}}
\end{minted}
\end{minipage}
\caption{\textbf{HLS code for three different implementations of GEMM kernels} ---
The loop unrolling factors are set as 32.
The latency, II, and frequency results are obtained from the HLS report.
}
\Description{HLS code for three different implementations of GEMM kernels.
The loop unrolling factors are set as 32.
The latency, II, and frequency results are obtained from the HLS report.}
\label{fig:single_hls}
\end{figure}
We still leverage the GEMM kernel as an example.
Even in this simple case, achieving high performance is not straightforward.

\paragraph{\textbf{Pitfall \RNum{1}: Simply inserting pragmas cannot lead to high performance.}}
As depicted on the left in Fig.~\ref{fig:single_hls}, an HLS programmer initially defines a vanilla floating-point GEMM kernel of size $1024\times 1024$, consisting of a loop nest of three levels.
If this code is directly fed to HLS, the resulting latency is \SI{25074}{ms} even though the HLS tool attempts to automatically pipeline the inner loop.

To further exploit the parallelism of the kernel, an intuitive idea is to unroll and pipeline the innermost loop.
Programmers can specify the target initiation interval (II) of the design using \mintinline{cpp}{#pragma pipeline}.
Given that the innermost loop is unrolled with a factor of 32, arrays \code{A} and \code{B} need to be partitioned into multiple banks to facilitate parallel access.
Surprisingly, the latency does not reduce to 1/32 but only 70\% of the latency of the original design, with an unfavorable increase in II.
This is primarily due to a loop-carried dependency in the floating-point accumulation of \code{C[i][j]}, which requires more than one cycle to finish, preventing effective pipelining with an II equal to one~\cite{johannes2021hls_tpds}.
Furthermore, the increased II leads to a reduced frequency, potentially causing routing issues during backend synthesis.

To resolve this issue, we can change the loop order to avoid updating the same matrix element in consecutive iterations.
As shown on the right of Fig.~\ref{fig:single_hls}, by swapping the loops of \code{j} and \code{k} (Lines 18-25), we transform the accumulation pattern into row-wise product, ensuring that adjacent iterations update different elements of the output matrix.
Additionally, a buffer of size 1024 is introduced to store intermediate results (Line 8), which are written back to memory only after iterating through one row.
As a result, we can achieve a \SI{112}{ms} latency with II=1, which achieves a 223$\times$ speedup compared to the vanilla implementation.

This example underscores the importance of source-level transformation in HLS-based hardware accelerator design.
Adding pragmas alone does not result in high performance; instead, it requires careful program restructuring to enable desired optimizations.
Unfortunately, even with the latest design-space exploration (DSE) techniques in HLS compilers~\cite{yehpca2022scalehls,zhao2017comba,atefeh2021autodse,atefeh2021gnndse}, identifying such optimizations may prove challenging.
These DSE methods commonly search for parameters associated with loop splitting, pipelining, or unrolling, yet they often lack support for crucial memory customizations, as discussed in \S\ref{sub:evaluation}.
\Name resolves this issue by providing memory customization primitives, allowing users to insert buffers at a given axis (\S\ref{sub:sch_constr}).
A full \Name example can be found in Supplementary Material \textcolor{ACMPurple}{C}.


Further optimizing a GEMM kernel may adopt a systolic array architecture, which requires streaming connections between multiple processing elements and constructing complex I/O networks to achieve high performance.
These optimizations require substantial code rewriting and also cannot be accomplished by simple pragma insertion.
For example, a high-performance 2$\times$2 systolic array for GEMM already requires more than 1,100 lines of C++ code~\cite{wang2021autosa}, which demonstrates the complexity of single-kernel HLS design.

\subsection{Multi-Kernel Design}
\label{sub:motivation_multi}
Once we have optimized a single-kernel GEMM design, the next challenge is to employ it as a fundamental building block for large designs.
In this context, we aim to construct a two-layer feed-forward network (FFN) module, a component commonly used in Transformer models \cite{devlin2018bert,radford2019gpt2,vaswani2017transformer}.
However, it remains a non-trivial task even though we already have an optimized GEMM kernel.

\begin{figure}[t]
\begin{minipage}{0.54\linewidth}
\begin{minted}[linenos,
               xleftmargin=1.8em,
               fontsize=\scriptsize,
               tabsize=2,
               escapeinside=||,
               autogobble]{cpp}
// Simple cascade
void rp_gemm(
  float A[1024][1024], float B[1024][1024],
  float C[1024][1024]
) { /*  See Fig. 2  */ }

void top(float X[1024][1024],
         float W_A[1024][1024],
         float W_B[1024][1024],
         float Y[1024][1024]) {
  float Z[1024][1024];
  rp_gemm(X, W_A, Z);
  rp_gemm(Z, W_B, Y);
}
\end{minted}
\bigskip
\resizebox{0.9\linewidth}{!}{
\scriptsize
\begin{tabular}{lccccc}\Xhline{1pt}
               & \thead{Latency (ms)} & \thead{BRAM} & \thead{DSP} &   \thead{FF}  &  \thead{LUT}  \\\Xhline{1pt}
Simple cascade &          280 & 1984 & \textcolor{red}{320} & 42761 & 24896 \\
+ \code{rp_gemm}      &          112 &   64 & 160 & 21391 & 11765 \\
\textcolor{red}{+ \code{rp_gemm_1}}    &          \textcolor{red}{168} &   64 & 160 & 21364 & 11856 \\\Xhline{1pt}
Interface uni.        &          \textbf{224} & 1920 & 160 & 21377 & 16068 \\
+ \code{rp_gemm}      &          112 &   64 & 160 & 21372 & 11913 \\\Xhline{1pt}
\end{tabular}
}
\end{minipage}
\begin{minipage}{0.45\linewidth}
\begin{minted}[linenos,
               fontsize=\scriptsize,
               escapeinside=||,
               autogobble]{cpp}
// Interface unification
void rp_gemm(
  float A[1024][1024], float B[1024][1024],
  float C[1024][1024]
) {
  // explicitly partition A
  #pragma partition |\colorbox{yellow}{var=A}| cyclic factor=32 dim=1
  #pragma partition var=B cyclic factor=32 dim=1
  #pragma partition var=C cyclic factor=32 dim=1
  // ...
}

void top(float X[1024][1024],
         float W_A[1024][1024],
         float W_B[1024][1024],
         float Y[1024][1024]) {
  #pragma allocation |\colorbox{yellow}{instances=rp\_gemm}| limit=1
  float Z[1024][1024];
  rp_gemm(X, W_A, Z);
  rp_gemm(Z, W_B, Y);
}
\end{minted}
\end{minipage}
\caption{\textbf{HLS code for cascading two GEMM kernels} --- Changes are highlighted in yellow.}
\Description{\textbf{HLS code for cascading two GEMM kernels} --- Changes are highlighted in yellow.}
\label{fig:multi_hls}
\end{figure}

\paragraph{\textbf{Pitfall \RNum{2}: Simply calling optimized kernels does not guarantee a high-quality design.}}
As depicted on the left of Fig.~\ref{fig:multi_hls}, within the top-level function, we input an initial tensor \code{X} and two weight parameters, \code{W_A} and \code{W_B}, followed by the output being written to \code{Y} (Lines 7-10).
In the main body, we create an intermediate tensor \code{Z} (Line 11), reuse the \code{rp_gemm} kernel defined in Fig.~\ref{fig:single_hls}, and invoke it twice to perform a linear layer computation (Lines 12-13).
This approach intuitively chains two function calls together.

Based on the results in Fig.~\ref{fig:single_hls}, cascading two GEMM kernels should yield a latency of \SI{224}{ms}, since a single-kernel GEMM has a latency of \SI{112}{ms}.
However, the HLS report in Fig.~\ref{fig:multi_hls} indicates a latency of \SI{280}{ms}, which is 1.25$\times$ slower than expected.
Furthermore, reusing the GEMM kernel for these two function calls should maintain resource usage at the same level as a single kernel, yet the HLS report indicates a doubling of resource utilization.
Closer examination reveals that HLS generates two distinct copies of the GEMM kernel, named \code{rp_gemm} and \code{rp_gemm_1}, with \code{rp_gemm_1} exhibiting a worse latency than \code{rp_gemm}.
The root cause is the function interface, where, in Fig.~\ref{fig:multi_hls}, we partition the second and third arguments (\code{A} and \code{B}) for the \code{rp_gemm} function. These two arguments correspond to the arrays \code{W_A} and \code{Z} in the top-level function.
However, \code{Z}, already a partitioned array, is once again passed into \code{rp_gemm} as the first argument, triggering partitioning of the first argument of the \code{rp_gemm} function.
This divergence in partitioning leads HLS to view the two \code{rp_gemm} kernels as distinct, with the first kernel partitioning the latter two arguments, while the second kernel partitions all three arguments.
Thus, two different copies of the \code{rp_gemm} kernel are generated.
Consequently, two distinct copies of the \code{rp_gemm} kernel are generated, and an unintended partition scheme causes HLS to make incorrect assumptions about loop variable dependencies, resulting in increased latency.

To rectify this issue and ensure proper sharing of function units while generating a design with the anticipated latency, we work towards unifying the function interface.
As shown on the right of Fig.~\ref{fig:multi_hls}, we explicitly partition the first argument of the \code{rp_gemm} kernel, thereby ensuring that all inputs and outputs are partitioned consistently.
Additionally, we enforce an \code{allocation} pragma to ensure the generation of only one function instance.
As a result, HLS produces a single copy of the \code{rp_gemm} kernel, as indicated by the resource usage in the bottom-left of Fig.~\ref{fig:multi_hls}.
Moreover, the latency is twice that of a single-kernel latency, totaling \SI{224}{ms}, aligning with our expectations.

This example highlights the inherent complexity of composing multiple kernels, requiring careful consideration of appropriate interfaces for each kernel and effective connection through intermediate buffers.
\Name introduces composable schedules and holistic optimizations to resolve this issue.
Further insights will be discussed in \S\ref{sec:composition}.

\section{\Name Overview}
\label{sec:design}

\begin{table}[t]
\caption{\textbf{Comparison between \Name and existing high-level hardware languages} --- APL denotes accelerator programming language, ADL is accelerator design language, and IL denotes intermediate language.}
\label{tab:comparison}
    \centering
    \resizebox{\linewidth}{!}{
    \begin{tabular}{lcccccccccccc}\Xhline{1pt}
         &  & & \multicolumn{8}{c}{\textbf{Single-Kernel Design (\S\ref{sec:customization})}} & \multicolumn{2}{c}{\textbf{Multi-Kernel Design (\S\ref{sec:composition})}}\\\cmidrule[1pt](lr){5-11}\cmidrule[1pt](lr){12-13}
         \multirow{2}{*}{\textbf{Framework}}&  \multirow{2}{*}{\textbf{Type}} & \textbf{Input}&  \textbf{Output}&  \textbf{Decoupled} & \multicolumn{4}{c}{\textbf{Customizations}}&  \textbf{Verifiable} & \textbf{Template} &  
\textbf{Composable} & \textbf{Dataflow}\\
        &  & \textbf{Language} & \textbf{Format} & \textbf{Y/N} & \textbf{Compute} & \textbf{Memory} & \textbf{Types} & \textbf{Comm.} & \textbf{Rewrites} & \textbf{Kernels} & \textbf{Optimizations} & \textbf{Optimizations}\\\Xhline{1pt}
         
         TVM~\cite{chen2018tvm}&  APL& Python &N/A&\gcmark&  \gcmark&  \gcmark&  \gcmark&  \rxmark& \rxmark& \gcmark &\rxmark & \rxmark
\\
         Exo~\cite{yuka2022exo}& APL& Python &N/A& \gcmark& \gcmark& \gcmark& \gcmark& \rxmark& \gcmark& \rxmark &\rxmark &\rxmark 
\\
        Spatial~\cite{koeplinger2018spatial}&  ADL&  Custom &Chisel& \rxmark&  \gcmark&  \gcmark&  \gcmark&  \gcmark&\rxmark & \gcmark &\rxmark &\gcmark
\\
        Aetherling~\cite{durst2020aetherling} &ADL & Haskell &Chisel& \rxmark & \rxmark & \gcmark& \rxmark &\rxmark &\rxmark &\rxmark &\rxmark &\gcmark
\\
        Fleet~\cite{thomas2020fleet} & ADL & Scala &Chisel& \rxmark &\rxmark &\gcmark &\rxmark &\rxmark & \rxmark&\rxmark &\rxmark &\rxmark
\\
        ScaleHLS~\cite{yehpca2022scalehls}&  IL& C++/\MLIR& HLS C++ &\rxmark&  \gcmark&  \rxmark&  \rxmark&  \rxmark& \rxmark& \rxmark &\rxmark &\rxmark
\\
        PyLog~\cite{huang2021pylog}&  ADL&  Python & HLS C++ &\rxmark&  \gcmark&  \gcmark&  \gcmark&  \rxmark& \rxmark& \rxmark &\rxmark &\rxmark
\\
        HeteroCL~\cite{lai2019heterocl}&  ADL & Python&  HLS C++ &\gcmark&  \gcmark&  \gcmark&  \gcmark&  \rxmark& \rxmark& \rxmark& \rxmark &\rxmark
\\
        Dahlia~\cite{rachit2020dahlia} &ADL & Custom & HLS C++&\rxmark &\gcmark &\gcmark &\rxmark &\rxmark & \rxmark&\rxmark &\rxmark & \rxmark
\\
 \textbf{Allo}& ADL/IL& Python/\MLIR & HLS C++ &\gcmark& \gcmark& \gcmark& \gcmark& \gcmark&\gcmark&\gcmark &\gcmark&\gcmark\\\Xhline{1pt}
    \end{tabular}
    }
\end{table}


Recently, various accelerator design languages (ADLs) have been proposed to mitigate the limitations of HLS.
Some of these approaches expose hardware customizations in a higher-level language, requiring users to follow specific coding styles and relying on compilers to generate high-performance implementations~\cite{koeplinger2018spatial,durst2020aetherling,thomas2020fleet}.
While this approach can partially resolve Pitfall \RNum{1} if the compiler is able to generate a proper memory hierarchy for the design, it requires users to write code in a functional language or in their custom formats, subsequently generating HDL code in Verilog or Chisel~\cite{bachrach2012chisel}.
This imposes a significant burden on programmers to translate their applications and debug in these languages.
Conversely, other ADLs are built on top of the original HLS C++ toolchain~\cite{lai2019heterocl,rachit2020dahlia,huang2021pylog}.
HeteroCL~\cite{lai2019heterocl} introduces the concept of separation of concerns in hardware design and provides primitives for users to optimize the program.
Dahlia~\cite{rachit2020dahlia} proposes a type system to ensure the consistency of memory partitioning and loop unrolling but lacks crucial customizations for pipelining and dataflow.
Both ScaleHLS~\cite{yehpca2022scalehls} and PyLog~\cite{huang2021pylog} automate hardware design and produce HLS C++ code as output.
Nevertheless, most of these ADLs focus on optimizing a single kernel and cannot efficiently address Pitfall \RNum{2}.

In this section, we present an overview of the \Name programming model and compilation flow.
A comparison between \Name and other high-level hardware languages and compilers is listed in Table~\ref{tab:comparison}.
\Name fully decouples hardware customizations from algorithm specifications, with a particular focus on enhancing memory and communication customizations.
This approach effectively addresses Pitfall \RNum{1} (see \S\ref{sec:customization}).
Furthermore, \Name provides the ability to declare parameterized kernels, thereby improving the usability of single-kernel designs.
What differentiates \Name from other ADLs is its ability to compose individual kernels and construct large-scale, high-performance designs.
\Name proposes composable schedules and holistic dataflow optimization to efficiently tackle Pitfall \RNum{2} (see \S\ref{sec:composition}).
Moreover, we leverage \Name to design a spatial architecture for large language models (LLMs) and execute the design on an FPGA.
The FPGA on-board evaluation shows its functionality and high performance, which is unachievable by prior ADLs (\S\ref{sub:on_board}).

\begin{figure}[!htbp]
    \centering
    \includegraphics[width=\linewidth]{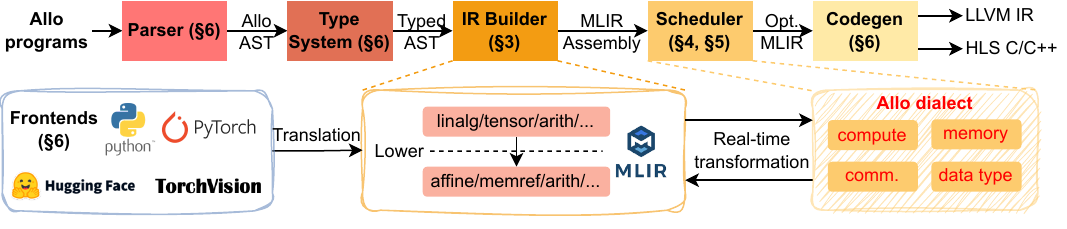}
    \caption{\textbf{Overview of the \Name compilation flow.}}
    \Description{Overview of the \Name compilation flow.}
    \label{fig:overview}
\end{figure}

\begin{table}[t]
    \centering
    \caption{\textbf{A partial list of the customization primitives supported by \Name.}}
    \label{tab:primitives}
    \resizebox{\linewidth}{!}{
    \begin{tabular}{lp{6cm}lp{6cm}}\Xhline{1pt}
 \multicolumn{2}{c}{\textbf{Compute Customizations}}& \multicolumn{2}{c}{\textbf{Memory Customizations}}\\\Xhline{1pt}
         \code{s.split(i,v)}&  Split loop \code{i} into a two-level nested loop with \code{v} as the bound of the inner loop. &  \code{s.buffer_at(A,i)}& Create an intermediate buffer at loop \code{i} to store the results of array \code{A}.
\\
         \code{s.fuse(*l)}&  Fuse multiple sub-loops \code{l} in the same nest loop into one. &  \code{s.reuse_at(A,i)}& Create a buffer storing the values of array \code{A}, where the values are reused at loop \code{i}.
\\
         \code{s.reorder(*l)}&  Switch the order of sub-loops \code{l} in the same nest loop. &  \code{s.partition(A,d,v)}& Cyclic/Block partition dimension \code{d} of array \code{A} with a factor \code{v}. 
\\
         \makecell[l]{\code{s.compute_at}\\\code{(Op1,Op2,i)}}& \makecell[l]{Merge loop \code{i} of the operation \code{Op1} to the\\ corresponding loop level in operation \code{Op2}.} &  \code{s.pack(A,i,v)}& Pack dimension \code{i} of array \code{A} into words with a factor \code{v}. 
\\
         \code{s.unroll(i,v)}&  Unroll loop \code{i} by factor \code{v}. 
\\\cmidrule[1pt]{3-4}
         \code{s.unfold(i)}&  Unfold loop \code{i} as hardware instances. &  \multicolumn{2}{c}{\textbf{Communication Customizations}}
\\\cmidrule[1pt]{3-4}
         \code{s.pipeline(i,v)} &  Schedule loop \code{i} in a pipeline manner with a target initiation interval \code{v}. &  \code{s.relay(A,Dst,v)}& Connect array \code{A} to destination \code{Dst} with a FIFO of depth \code{v}.\\\Xhline{1pt}
    \end{tabular}
    }
    \vspace{-.1in}
\end{table}

As illustrated in Fig.~\ref{fig:overview}, \Name offers a Python-embedded ADL for improved productivity, and the \Name compiler follows a conventional compilation workflow.
Users can either use the Python frontend to write a Python kernel or leverage the PyTorch frontend to directly import deep learning models from TorchVision~\cite{torchvision2016} and HuggingFace~\cite{wolf2019huggingface}, which will be further discussed in \S\ref{sec:impl}.
For the Python kernel, \Name first parses it into an abstract syntax tree (AST).
The AST then proceeds through a type system that performs essential tasks such as type checking, type inference, and type conversions based on user-provided annotations (\S\ref{sec:impl}).
The typed AST is subsequently passed into an intermediate representation (IR) builder.
We develop an \Name dialect within the \MLIR ecosystem, which facilitates the separation of hardware customizations at the IR level (\S\ref{sec:customization}).
The IR builder generates \MLIR programs for customization and code generation.
Once the IR is obtained, program transformations are applied using the provided customization primitives (\S\ref{sec:customization}).
This approach seamlessly integrates with different in-tree \MLIR dialects, enabling the generation of high-performance designs for diverse backend targets.
Lastly, we generate \LLVM IR~\cite{chris2004llvm} for CPU simulation and HLS C/C++~\cite{vitis_hls_2022} for hardware synthesis (\S\ref{sec:impl}).
Notice most of the hardware customizations are target-independent, allowing our compilation flow to target ASIC designs as well.
We plan to integrate the CIRCT~\cite{CIRCT} project as our backend to support custom circuit generation.

\section{Customizable Hardware Transformations}
\label{sec:customization}
In this section, we use a systolic array~\cite{kung1978systolic} as an example to illustrate the language features of \Name, showcasing its capabilities in handling complex transformations for a single kernel design.
The systolic array is a prevalent spatial architecture extensively employed in deep learning accelerators such as Google TPUs~\cite{jouppi2017tpuv1} and AWS Inferentia~\cite{aws_inferentia}.
It comprises a set of processing elements (PEs) that iteratively execute repetitive operations.
By reusing data across these PEs, it minimizes off-chip memory access, resulting in high performance with minimal energy consumption.

\subsection{Schedule Construction}
\label{sub:sch_constr}
\begin{figure}[t]
\begin{minipage}{0.45\linewidth}
\begin{minted}[linenos,
               fontsize=\scriptsize,
               escapeinside=||]{python}
# Algorithm specification
def gemm(A: int8[M, K], B: int8[K, N],
         C: int16[M, N]):
  for i, j in allo.grid(M, N, "PE"):
    for k in range(K):
      C[i, j] += A[i, k] * B[k, j]

# Schedule construction
s = allo.customize(gemm)
buf_A = s.buffer_at(s.A, "j")            # |$p_0$|
buf_B = s.buffer_at(s.B, "j")            # |$p_1$|
pe = s.unfold("PE", axis=[0, 1])         # |$p_2$|
s.partition(s.C, dim=[0, 1])             # |$p_3$|
s.partition(s.A, dim=0)                  # |$p_4$|
s.partition(s.B, dim=1)                  # |$p_5$|
s.relay(buf_A, pe, axis=1, depth=M + 1)  # |$p_6$|
s.relay(buf_B, pe, axis=0, depth=N + 1)  # |$p_7$|
\end{minted}
\end{minipage}
\begin{minipage}{0.42\linewidth}
\begin{center}
\includegraphics[width=0.9\linewidth]{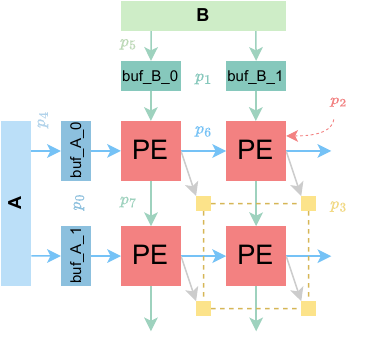}
\end{center}
\end{minipage}
\caption{\textbf{\Name program for an integer output-stationary systolic array} ---
\code{M}, \code{N}, and \code{K} are predefined integer constants.
\code{allo.grid} is a syntactic sugar for multiple Python \code{range}-for loops.
\code{A} and \code{B} are located in DRAM.
}
\Description{Code block for \Name program for a systolic array.}
\label{fig:systolic_schedule}
\end{figure}

Given the initial algorithm definition in Lines 2-6 of Fig.~\ref{fig:systolic_schedule}, we transform the algorithm specification into a tangible hardware implementation.
Here, we formally define a \emph{schedule} $\mathcal{S}$ of a program $P_0$ as a sequence of transformations $(p_i)_{i=1}^N$ such that
\begin{equation}
P_0\overset{p_1}{\rightsquigarrow} P_1\overset{p_2}{\rightsquigarrow}\cdots\overset{p_N}{\rightsquigarrow}P_N\,,
\end{equation}
where $\overset{p_i}{\rightsquigarrow}$ denotes a program rewrite with a primitive $p_i$, and $N$ is the number of customization primitives in this schedule.

Table~\ref{tab:primitives} lists the primitives supported by \Name.
The compute customizations transform the loops and attach necessary attributes, which inherit the idea from existing schedule languages~\cite{jrk2013halide,chen2018tvm,lai2019heterocl,yuka2022exo}.
Notice instead of implementing monolithic compiler passes for the primitives~\cite{jrk2013halide,chen2018tvm,lai2019heterocl}, \Name adopts an approach akin to Exo~\cite{yuka2022exo}, which treats primitives as program rewrites, ensuring correctness for each transformation.
Users can print the intermediate module after each customization to inspect real-time program transformations, providing deeper insights into the customization primitives. 
Moreover, we develop an \Name-\MLIR dialect to implement those customizations primitives at the IR level, with each primitive corresponding to an operation in the \Name dialect.
It enables \Name to serve as an intermediate language and support different frontends.
In the subsequent discussion, we will primarily focus on memory and communication customizations, which distinguish \Name from other ADLs.

As illustrated in Fig.~\ref{fig:systolic_schedule}, users can create a schedule by invoking the \code{allo.customize} function and progressively apply primitives to the newly-formed schedule (Line 9).
Each primitive exactly does one transformation as shown on the right of Fig.~\ref{fig:systolic_schedule}.
We start by creating intermediate buffers for \code{A} and \code{B} arrays (Lines 10-11), which creates a line buffer for peripheral PEs to efficiently load data from off-chip memory.
After that, we can easily generate $M\times N$ PEs along the 0th and 1st axes by invoking \code{.unfold()} (Line 12).
Further configurations on the number of PEs can be controlled by loop tiling using the \code{.split()} primitive.
As these PEs are actual hardware instances, and each PE requires parallel memory access, \code{C} is partitioned in both dimensions to accommodate the nature of output-stationary accumulation (Line 13).
Furthermore, \code{A} and \code{B} arrays are also partitioned to create multiple banks, facilitating parallel data input into the line buffers (Lines 14-15).
More importantly, users need to specify the intra-kernel communication.
\Name offers a seamless way to connect neighboring PEs with FIFOs through the \code{.relay()} primitive.
As shown in Lines 16-17, \code{buf_A} is connected along the 1st axis, while \code{buf_B} is connected along the 0th axis, and these connections will subsequently be synthesized as FIFOs within the hardware.
A generated HLS C++ code for this systolic array is attached in Supplementary Material \textcolor{ACMPurple}{D} for reference.

Notice this approach is general enough to allow users to express programs in different forms.
Users can start from an arbitrary program status $P_i$ using our programming model and apply customization primitives $p_i$ to obtain a transformed program $P_{i+1}$.
For example, in Fig.~\ref{fig:systolic_schedule}, the program $P_1$ after applying $p_0$ is still functional, as it essentially moves data from DRAM to an intermediate buffer and lets the computation logic load the same data from this buffer.
This is not achievable by ad-hoc systolic array compilers like AutoSA~\cite{wang2021autosa}.
The optimization process in AutoSA is not transparent, and programmers cannot easily configure the architecture of the generated systolic arrays.
In contrast, our approach allows customization of compute, memory, data types, and communication.
We can further create an additional memory hierarchy using the \code{.buffer_at()} primitive for \code{buf_A} and \code{buf_B}, which reduces the memory fan-out to one.
Additionally, \Name provides more flexibility than semi-manual systolic array generators like SuSy~\cite{lai2020susy} and T2S~\cite{nitish2019t2s}, which require users to write verbose uniform recurrence equations and conduct complex space-time transformations manually.
On the contrary, \Name can start from a vanilla GEMM kernel and progressively transform it into a functional systolic array using eight lines of schedule code.
Leveraging the provided primitives, \Name can strike the right balance between compiler-based optimizations and manual optimizations.
As demonstrated in \S\ref{sub:evaluation}, our programming model can not only be used to generate systolic arrays but is also general enough to support different applications.

\subsection{Verification}
\label{sub:verification}
\begin{figure}[t]
    \begin{subfigure}[b]{0.38\textwidth}
        \label{fig:verif-allo}
        \begin{minipage}{0.38\textwidth}
        \begin{minted}[linenos, fontsize=\scriptsize, escapeinside=||]{python}
# Algorithm specification
def gemm1(A: float[M, K], B: float[K, N],
          C: float[M, N]):
    ...
def gemm2(C: float[M, N], D: float[N, P],
          E: float[M, P]):
    ...
def two_mm(A: float[M, K], B: float[K, N],
           D: float[N, P], E: float[M, P]):
    C: float[M, N]
    gemm1(A, B, C)
    gemm2(C, D, E)
# Schedule construction
s_orig = allo.customize(two_mm)
# Duplicate schedule for verification
s = allo.customize(two_mm)
s.reorder("gemm2:k", "gemm2:j")
s.buffer_at(s.C, axis="gemm2:i")
s.relay(s.A, "gemm2")
|\textcolor{red}{s.reorder(}|"gemm1:j"|\textcolor{red}{,}| "gemm1:i"|\textcolor{red}{)}|
# 1. Functional simulation testing
f = s.build()
# ... Initialize NumPy arrays (omitted)
f(np_A, np_B, np_D, np_E)
# 2. Formal equivalence checking
allo.verify(s, s_orig)
        \end{minted}
        \end{minipage}
    \caption{\Name code snippet}
    \label{subfig:twomm}
    \end{subfigure}
    \hspace{.35in}
    \begin{subfigure}[b]{0.45\textwidth}
        \begin{minipage}{0.45\textwidth}
        \begin{minted}[linenos, fontsize=\scriptsize, escapeinside=||]{cpp}
void gemm1(float A[M][K], float B[K][N], stream C) {
  |\color{red}{for (int j = 0; j < N; j++)}| {
    |\color{red}{for (int i = 0; i < M; i++)}| {
      float sum = 0;
      for (int k = 0; k < K; k++) { 
        sum += A[i][k] * B[k][j];
      C.write(sum);
}}}}
void gemm2(stream C, float D[N][P], float E[M][P]) {
  float buf_E[N];
  for (int i = 0; j < M; j++) {
    // ... Initialize buf_E (omitted)
    for (int k = 0; k < N; k++) {
      // A mismatch between read and write
      float c = C.read();
      for (int j = 0; j < P; j++) {
        buf_E[j] += c * D[k][j];
    }}
    // ... Write-back to E (omitted)
}}
void two_mm(float A[M][K], float B[K][N], 
            float D[N][P], float E[M][P]) {
  stream C_fifo;
  gemm1(A, B, C_fifo);
  gemm2(C_fifo, D, E);
}
        \end{minted}
        \end{minipage}
    \caption{Corresponding HLS C++ code snippet}
    \label{subfig:twomm_c}
    \end{subfigure}
\caption{\textbf{A buggy \Name example with data streaming and loop reordering} --- The two matrix multiplications are computed back-to-back. The code marked in \textcolor{red}{red} indicates bugs in the program.}
\label{fig:verif}
\Description{A buggy \Name example with data streaming and loop reordering.}
\end{figure}
Ensuring the correctness of the generated accelerator is of great importance.
\Name employs two key verification procedures to enhance the reliability of the generated code. First, \Name leverages the CPU backend to conduct functional simulation testing (\S\ref{sec:impl}). Second, \Name integrates an equivalence checker~\cite{pouchet2024formal} to formally verify the equivalence of the programs before and after customizations, provided that the programs have statically interpretable control-flow (SICF). SICF requires the problem size to be known at compile-time and does not support parametric loop nest analysis. 

Fig.~\ref{fig:verif} shows a data access order bug caused by incorrect customizations. In this example, Lines 17-18 of Fig.~\ref{subfig:twomm} transform the second matrix multiplication from an inner-product to a row-wise product, which reads the input \code{C} in a row-major order. However, Line 20 reorders the first matrix multiplication loops to send the output \code{C} in a column-major order. This discrepancy in data receiving and sending orders violates the requirement of in-order access on a stream FIFO. The \texttt{.verify()} on Line 26 invokes the equivalence checker which takes the schedule before (Line 14) and after customizations (Line 20) to formally verify the program semantic equivalence.
In this example, the customizations break the accelerator design and cause a semantic difference in the customized code.
The difference in the symbolic representation is detected and reported by the equivalence checker to facilitate debugging.
Notice the verification can be conducted after each primitive is applied, ensuring the correctness of the transformations at each step.

\subsection{Parameterized Kernel Templates}
We initially constructed a systolic array with fixed dimensions and data types, which lacks flexibility when handling variable-sized input matrices.
In the following, we leverage the previously defined systolic array to introduce a tiled design that accommodates inputs of arbitrary sizes.

\begin{figure}[!htbp]
\begin{minipage}{0.7\linewidth}
\begin{minted}[linenos,
               fontsize=\scriptsize]{python}
def systolic[TyA, TyB, TyC, Mt: index, Nt: index, K: index]
    (A: TyA[Mt, K], B: TyB[K, Nt], C: TyC[Mt, Nt])

def tiled_systolic[TyA, TyB, TyC, M: index, N: index, K: index]
    (A: TyA[M, K], B: TyB[K, N], C: TyC[M, N]):
    local_A: TyA[8, K]; local_B: TyB[K, 8]; local_C: TyC[8, 8]
    for mi, ni in allo.grid(M // 8, N // 8, name="outer_tile"):
        # ... load_A_tile, load_B_tile
        systolic[TyA, TyB, TyC, 8, 8, K](local_A, local_B, local_C)
        # ... store_C_tile
\end{minted}
\end{minipage}
\caption{\textbf{Tiled systolic array implementation in \Name.}}
\Description{Tiled systolic array implementation in \Name.}
\label{fig:tiled_systolic}
\end{figure}

\Name provides a user-friendly parameterization template to facilitate polymorphism.
As illustrated in Fig.~\ref{fig:tiled_systolic}, we parameterize the \code{systolic} function with type parameters.
Users can define the function signature using the syntax \code{def <func>[<type params>](<args>)}.
Again, given \Name's decoupling of data types from the algorithm specification, both data types and shapes can serve as type parameters.
\Name permits additional constraints for parameterized data types.
For instance, \code{Ty: (int32, float32)} specifies that the data type \code{Ty} must be either \code{int32} or \code{float32}.
If any other data types are used, an error is raised.
Within the \code{tiled_systolic} function, we partially specialize the inner systolic array with a fixed size of $8\times 8$, allowing us to derive a tiled version of the systolic array capable of accommodating inputs of varying dimensions.

In \Name, we have encapsulated several template kernels as libraries, each accompanied by predefined schedules.
These templates include commonly used deep learning operators and high-performance systolic arrays for matrix-matrix multiplications and matrix-vector multiplication.
This approach allows users to conveniently reuse these kernels for their own workloads, reducing the burden of writing efficient schedules.
Additionally, this parameterized interface facilitates auto-tuning and auto-scheduling~\cite{shao2022metaschedule,chen2018autotvm,zheng2020ansor}, which we plan to explore in future work.

\section{Composable Schedules}
\label{sec:composition}
In this section, we explore the process of composing multiple schedules to construct a complete design.
We begin by introducing the \code{.compose()} primitive and delve into the implementation details of the hierarchical dataflow graph.
We then present the algorithm for schedule replay and memory layout composition.
Lastly, we extend \Name to support the composition of external kernels and present an algorithm for holistic optimization.

\subsection{\code{.compose()} Primitive}
\begin{figure}[!htbp]
\begin{minipage}{0.4\linewidth}
\begin{minted}[linenos,
               fontsize=\scriptsize,
               xleftmargin=1.8em,
               autogobble]{python}
def top(X: int8[32, 64]) -> int8[64, 32]:
    Z: int8[32, 64] = 0
    Y: int8[64, 32] = 0
    W_A: int8[64, 64] = W_A_cst
    W_B: int8[64, 64] = W_B_cst
    tiled_systolic[int8, int8, int16, \
        32, 64, 64, "FFN1"](X, W_A, Z)
    tiled_systolic[int8, int8, int16, \
        64, 32, 64, "FFN2"](Z, W_B, Y)
    return Y

s_top = allo.customize(top)
s_top.compose(s) # `s` is an optimized schedule
s_top.relay(s_top.Z, "tiled_systolic_FFN2")
\end{minted}
\end{minipage}
\hfill
\begin{minipage}{0.55\linewidth}
\begin{center}
\includegraphics[width=\linewidth]{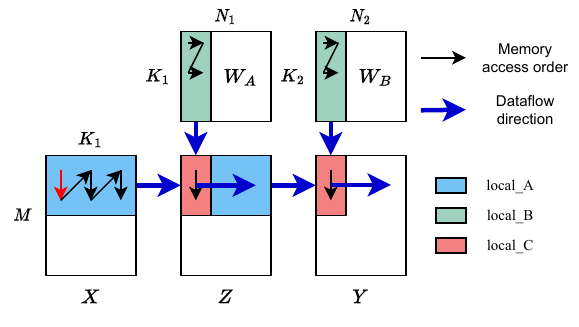}
\end{center}
\end{minipage}
\Description{Cascading two systolic arrays.}
\caption{\textbf{Cascading two systolic arrays} ---
\code{W_A_cst} and \code{W_B_cst} are predefined weight parameters.
Automatic type conversion on \code{Z} is performed between two kernels.}
\label{fig:multisystolic}
\end{figure}

We leverage the systolic array implementation in Fig.~\ref{fig:tiled_systolic} to illustrate how to cascade two systolic arrays to create a larger design.
As depicted in Fig.~\ref{fig:multisystolic}, the two systolic arrays are arranged in sequence, with an intermediate tensor \code{Z} facilitating data transfer.
This aligns with common practices in nowadays neural network implementations~\cite{paszke2019pytorch}.

Based on the primitives proposed in \S\ref{fig:overview}, we can optimize individual kernels independently.
However, we need a mechanism to connect these smaller kernels.
Therefore, we propose a new primitive, \code{s.compose(<new_schedule>,<id>)}, for incorporating a \code{<new_schedule>} into the original schedule \code{s}.
In Fig.~\ref{fig:multisystolic}, we call \code{s_top.compose(s)} to integrate an optimized schedule \code{s} of \code{tiled_systolic} to \code{s_top} (Line 13).
The additional \code{<id>} argument is used to distinguish between different function calls.
For instance, there are two callers to \code{tiled_systolic} in Lines 6 and 8.
By including an additional identifier in the type parameter list (e.g., \code{"FFN1"}, \code{"FFN2"}), users can customize them differently.
The \Name compiler will generate two distinct instances for those functions.
This fine-grained control empowers users to customize specific functions to meet their requirements, offering flexibility not available in traditional compilers with fixed patterns in the compiler pass.
Finally, the two systolic arrays are linked with another FIFO to establish a dataflow at the top level (Line 14).

Moreover, \code{.compose()} primitive can act as an IP integrator to import C++ kernels from existing HLS-based kernel libraries~\cite{vitis_hls_library,vitis_ai}.
\Name parses the function interface of the HLS IPs, and the function arguments will be translated into internal representations labeled with \Name-supported data types.
As long as users specify the correct number of arguments, \Name automatically generates wrapper functions with PyBind11~\cite{pybind11}, making it more general and extensible to different customizations.

\subsection{Hierarchical Dataflow Graph}
\label{sub:dfg}
\begin{figure}[!htbp]
\begin{minipage}{0.45\linewidth}
\begin{minted}[linenos,
               fontsize=\scriptsize]{python}
def add(A: T[M, N]) -> T[M, N]:
    B = A + 1
    return B
def mul(A: T[M, K], B: T[K, N]) -> T[M, N]:
    # ... Calculate matrix multiply of A and B
def top(A: T[M, K], B: T[K, N]) -> T[M, N]:
    C = mul(A, B)
    D = add(C)
    return D
\end{minted}
\end{minipage}
\begin{minipage}{0.4\linewidth}
\begin{center}
\includegraphics[width=0.85\linewidth]{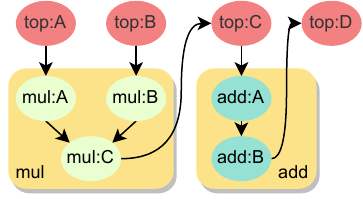}
\end{center}
\end{minipage}
\caption{\textbf{Example of a hierarchical dataflow graph.}}
\Description{Example of a hierarchical dataflow graph.}
\label{fig:dfg}
\end{figure}
Traditional dataflow graphs used in compilers for analysis are typically flattened, which removes the boundaries between functions and may miss potential optimization opportunities at the graph level~\cite{chen2018tvm,lai2019heterocl,reed2022fx}.
To preserve the hierarchy of the modules during scheduling, we require a new data structure capable of representing the dataflow graph in a way that facilitates the analysis of interfaces between functions.
As a solution, we propose a hierarchical dataflow graph that connects nested functions in a hierarchical way.

Given that the constructed \MLIR is in SSA form, each node in the dataflow graph represents an operation in the IR.
As shown in Fig.~\ref{fig:dfg}, the \code{top} function calls two subfunctions, \code{mul} and \code{add}, within its body.
To represent nodes in the dataflow graph, we use variable names with the function name as a prefix.
We maintain information about the caller and callee in the dataflow graph and explicitly connect function arguments with the caller's operands using edges.
For example, array \code{C} is passed into the function \code{add}, so there is an edge from \code{top:C} to \code{add:A}.
Since our primary concern is how data flows and not how many iterations are needed, control flow is eliminated from the graph.
This simplifies the composition process discussed in Section~\ref{sub:partition}.

\subsection{Schedule Replay}
We formalize the process of schedule composition as follows.
Consider a schedule of program $P$, represented as a sequence of primitives $\mathcal{S}_P=(p_0^{P};\ldots;p_{N_P}^P)$.
Composing this schedule $\mathcal{S}_P$ with another schedule $\mathcal{S}_Q$ involves appending the customization primitives of $\mathcal{S}_Q$ after the primitives of $\mathcal{S}_P$, i.e., $\mathcal{S}_Q\circ\mathcal{S}_P$, where $\circ$ denotes sequence concatenation.

We outline a general algorithm for composing two schedules in Algorithm~\ref{alg:sch_replay}, and it is easy to extend it to multiple schedules.
This algorithm traverses the primitives in the input schedules and replays them in the context of the new program space.
Since some functions may be renamed or duplicated due to the \code{<id>} interface, we must update the arguments of the primitives to ensure they apply to the correct function and operations (Line 3).
Before applying a new primitive, we verify that it will not conflict with previously applied primitives (Line 4).
For compute customizations, conflicts occur only when the same operation is targeted.
In such cases, an error is raised because the operation has already been transformed (Line 5).
Conflicts related to \code{.partition()} and \code{.relay()} will be discussed further in \S\ref{sub:partition}.
The primitive is then appended to the new schedule (Line 6).
Primitives are applied to the program $P$ only when all the sub-schedules are integrated into the top-level function and are ready for backend executable construction (Line 7), which saves redundant transformation time.
The resulting $\mathcal{S}_{out}$ can be used for subsequent composition.
This progressive composition process allows us to combine small designs step by step, culminating in the construction of a large design, with each submodule thoroughly tested, as discussed in \S\ref{sub:verification}.

\setlength{\textfloatsep}{0pt}
\begin{algorithm}[t]
\caption{\textbf{Composing multiple schedules with schedule replay}}
\label{alg:sch_replay}
\SetKwComment{Comment}{/* }{ */}
\KwData{Two schedules $\mathcal{S}_P$ and $\mathcal{S}_Q$ for programs $P$ and $Q$}
\KwResult{Composition of the schedules $\mathcal{S}_{out}=\mathcal{S}_Q\circ\mathcal{S}_P$ and the output program $P'$ after applying $\mathcal{S}_{out}$}
Initialize $\mathcal{S}_{out}=\mathcal{S}_P$\;
\ForEach{primitive $p_i\in\mathcal{S}_Q$}{
Update the arguments of $p_i$ to refer to the functions and arguments in program $P$\;
\If{$p_i$ conflicts with primitives in $\mathcal{S}_{out}$}{
Composition fails, raise an error\;
}
Append $p_i$ to $\mathcal{S}_{out}$\;
}
Apply each primitive in $\mathcal{S}_{out}$ to the program $P$ to obtain $P'$
\end{algorithm}
\setlength{\textfloatsep}{\textfloatsepsave}

\subsection{Memory Layout Composition}
\label{sub:partition}
When the customization primitives only affect the inner computation, it is straightforward to replay them with Algorithm~\ref{alg:sch_replay}.
However, complexity arises when schedules overlap through the function interface.
If a sub-schedule changes the function interface, the parent program must also change to avoid conflicts, as discussed in \S~\ref{sub:motivation_multi}.
It is important to maintain the consistency between function call arguments and actual function definitions.
Array partitioning is an example of this challenge.
In Fig.~\ref{fig:systolic_schedule}, when the schedule is integrated into Fig.~\ref{fig:tiled_systolic}, the \code{local_A}, \code{local_B}, and \code{local_C} arrays should also be partitioned.
This is because these arrays are partitioned within the \code{systolic} function and are passed into the function as arguments.

\begin{figure}[!htbp]
\begin{minipage}{0.4\linewidth}
\[\begin{aligned}
    \tau &:= (\hat{\tau}_1,\ldots,\hat{\tau}_N)\qquad\\
    \alpha &:= \mathbb{N}\\
    \hat{\tau} &:=\bot\mid\mathcal{C}_{\alpha}\mid\mathcal{B}_{\alpha}\mid\top
\end{aligned}\]
\end{minipage}
\begin{minipage}{0.48\linewidth}
    \includegraphics[width=\linewidth]{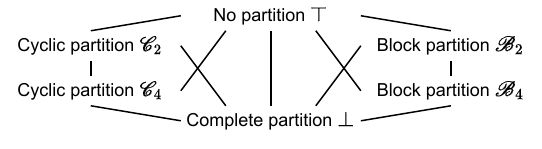}
\end{minipage}
\Description{Left: \textbf{Definition of the partition types.} Right: \textbf{Example lattice of partition types for a 1D array of shape $(8,)$} --- Since $8$ can be divided into $2\times 4$, there are six elements in this lattice.}
\caption{Left: \textbf{Definition of the partition types.} Right: \textbf{Example lattice of partition types for a 1D array of shape $(8,)$} --- Since $8$ can be divided into $2\times 4$, there are six elements in this lattice.}
\label{fig:lattice}
\end{figure}

Since hardware memory partitioning essentially alters data layout, 
we can explicitly represent data layouts as types~\cite{mangpo2014chlorophyll,chris2021mlir} and conduct analysis within this type system.
As shown in the left side of Fig.~\ref{fig:lattice}, we consider the partition type of an $N$-dimensional array.
The partition type $\tau$ is a composite type consisting of the base type $\hat{\tau}$ for each dimension.
Each base type can assume one of four choices.
$\bot$ means fully partitioning this dimension, allowing parallel access to all elements.
$\top$ represents no partition in this dimension, resulting in only one memory bank on the hardware.
$\mathcal{C}_\alpha$ represents cyclic partitioning with a factor of $\alpha$, 
where the elements in the original array are interleaved.
$\mathcal{B}_\alpha$ denotes block partitioning with a factor of $\alpha$, where the original array is divided into consecutive blocks.
Denote $s_i$ as the size of dimension $i$, and $\alpha$ should be an integer factor of $s_i$ (not including $1$ and itself).
We can construct subtype relations for these base types.
If $X<:Y$, it means the code expecting a memory with partition type $Y$ is also compatible with a memory with the partition type $X$.
For example, $\bot$ is a subtype of $\mathcal{C}_2$ because complete partitioning already partitions the array into cyclic with a factor of $2$.
If a kernel requires an array to be cyclic partitioned into two banks to access the elements in parallel, it is also fine to pass in a fully partitioned array since it offers more memory banks.
Notably, this subtyping relation is covariant, which means that the subtyping relation of base types $\hat{\tau}$ applies to composite types $\tau$ as well.
This subtyping relation actually forms a lattice, where each pair of elements in the type definition has a unique supremum $\top$ and a unique infimum $\bot$.
The right side of Fig.~\ref{fig:lattice} shows an example Hasse diagram.

After constructing this simple language for memory layout, we can define the typing rules for these partition types.
As shown in Fig.~\ref{fig:typing_memory}, the first row demonstrates the subtyping relations of base types, and the second row shows the composite rule and function application.
For function application, it is important to ensure the partition types of function signatures and caller operands are compatible, which results in a subtyping relation between $\tau_3$ and $\tau_1$.
For example, in Fig.~\ref{fig:multi_hls}, the original function \code{rp_gemm} has already partitioned array \code{Z} in $\mathcal{C}_{32}$, while the program attempts to pass in \code{Z}, which has no partitions (i.e., $\top$).
Therefore, our type system directly rejects this program, avoiding possible performance issues in Pitfall \RNum{2}.

\begin{figure}[!htbp]
\begin{mathpar}
\infer[S-Bottom-C]{\qquad}{\bot<:\mathcal{C}_\alpha}\qquad
\infer[S-Bottom-B]{\qquad}{\bot<:\mathcal{B}_\alpha}\qquad
\infer[S-Cyclic]{\alpha_2\equiv 0(\mathrm{mod}\,\alpha_1)}{\mathcal{C}_{\alpha_2}<:\mathcal{C}_{\alpha_1}}\qquad
\infer[S-Block]{\alpha_2\equiv 0(\mathrm{mod}\,\alpha_1)}{\mathcal{B}_{\alpha_2}<:\mathcal{B}_{\alpha_1}}\qquad
\infer[S-Top-C]{\qquad}{\mathcal{C}_\alpha<:\top}\qquad
\infer[S-Top-B]{\qquad}{\mathcal{B}_\alpha<:\top}\\
\infer[S-Array]{\exists i\in\{1,\ldots,N\}:\;\hat{\tau}_i<:\hat{\tau}_i'}{(\hat{\tau}_1,\ldots,\hat{\tau}_N)<:(\hat{\tau}_1',\ldots,\hat{\tau}_N')}\qquad
\infer[FuncApp]{\Gamma\vdash f:\tau_1\to\tau_2\quad \Gamma\vdash e:\tau_3\quad \tau_3<:\tau_1}{\Gamma\vdash f e:\tau_2}
\end{mathpar}
\caption{\textbf{A partial list of typing rules for the partition types in Fig.~\ref{fig:lattice}} --- $\Gamma$ is the typing context.}
\Description{\textbf{A partial list of typing rules for the partition types in Fig.~\ref{fig:lattice}} --- $\Gamma$ is the typing context.}
\label{fig:typing_memory}
\end{figure}

Based on the typing rules, new partition types need to be assigned to each variable in the program after schedule composition.
The unification algorithm commonly employed in functional languages for type inference does not fit in this case, due to the presence of subtyping relations~\cite{milner1982hm,hindley1969hm,damas1984typeinfer}.
In general, conducting type inference with subtypes can be a challenging task, which requires complex algebraic operations or leveraging an SMT solver to solve the constraints~\cite{dolan2017mlsub,lionel2020pearl,pottier1998subtype}.
However, given the lattice property of the subtyping relations, as well as the hierarchical dataflow graph constructed in \S\ref{sub:dfg}, we can apply dataflow analysis on this dataflow graph to efficiently assign types for the variables in the transformed program.

Consider a dataflow graph with $M$ nodes, we can use Algorithm~\ref{alg:layout} to calculate the proper memory layout of each node.
This iterative algorithm resembles the Worklist algorithm used in static dataflow analysis~\cite{kildall1973datflow}.
We use a concrete example in Fig.~\ref{fig:dfg} to illustrate the process.
Suppose we apply fully partition on array \code{C}, i.e., $t_{in}'=\bot$.
We first add the target node \code{top:C} and the target partition type $t_{in}'$ to the worklist (Line 1).
In the first iteration, we calculate \code{top:C}'s type as $t_{\text{\code{top:C}}}\gets \bot\sqcap\top=\bot$ (Line 4), where $\sqcap$ is the greatest lower bound (GLB) operator.
Since its type changes (Line 5), we traverse its predecessors (i.e., \code{mul:C}) and successors (i.e., \code{add:A}) (Line 6) and append them to the worklist since they are not in the same function (\code{top}) with \code{C} (Line 7-8).
Similarly, in the latter iterations, $t_{\text{\code{mul:C}}}$ and $t_{\text{\code{add:A}}}$ are updated to have type $\bot$, and no more dataflow nodes in the worklist update the types, which finalizes the algorithm.
Notice the $\sqcap$ operator in Line 4 is used to handle more general cases of merging two partition types, which makes sure, for example, even for two different types $\mathcal{C}_4$ and $\mathcal{B}_2$ in Fig.~\ref{fig:lattice}, they can be cast to a common type (e.g., $\bot$).

\setlength{\textfloatsep}{0pt}
\begin{algorithm}[t]
\caption{\textbf{Partition type inference (Memory layout propagation)}}
\label{alg:layout}
\KwData{The partition type $(t_1^{(0)},\ldots t_M^{(0)})$ of the nodes $(n_1,\ldots,n_M)$ in the hierarchical dataflow graph, and a \code{.partition()} primitive on node $n_{in}$ that transforms type $t_{in}$ to $t_{in}'$}
\KwResult{Result partition type $(t_1^{(out)},\ldots t_M^{(out)})$}
Initialize Worklist $\gets$ $\{(n_{in},t_{in}')\}$\;
\While{Worklist is not empty}{
Pick an item of dataflow node and target type $(n,t')$ from Worklist\;
Update type $t_{n}^{(next)}\gets t'\sqcap t_{n}^{(curr)}$\;
\If{$t_{n}^{(next)}\neq t_{n}^{(curr)}$}{
\ForEach{predecessors and successors $\tilde{n}$ of $n$}{
\If{$\tilde{n}$ and $n$ are in different functions}{
Add $(\tilde{n},t_{n}^{(next)})$ to Worklist\;
}}}}
\end{algorithm}
\setlength{\textfloatsep}{\textfloatsepsave}

Algorithm~\ref{alg:layout} is guaranteed to terminate in linear time, as formally stated in the following:
\begin{theorem}
\label{theorem:fixed_point}
Algorithm~\ref{alg:layout} can terminate in $\mathcal{O}(M)$ steps.
\end{theorem}

This termination condition can be established through the Knaster-Tarski Fixed-Point Theorem~\cite{tarski1955fixedpoint} given the fact that the depths of the lattices are fixed constants not related to $M$.
A formal proof can be found in Supplementary Material \textcolor{ACMPurple}{B}.
Since in each iteration, $t_i$ only changes from one partition type to another on the lattice in one direction, this algorithm is efficient in inferring the partition types.
Experimental results in \S\ref{sub:evaluation} demonstrate its overhead is negligible.

It is worth noting that Algorithm~\ref{alg:layout} can also be applied for streaming type propagation.
Due to space constraints, we do not provide the full details here.


\subsection{Holistic Optimization}
\label{sub:fifo_sizing}
To achieve high-performance spatial architecture, functions are interconnected using FIFOs, forming distinct dataflow stages.
While this creates a functional architecture, dataflow may suffer from performance issues, particularly when there are data stalls.
HLS alone cannot determine the ideal FIFO size between stages.
Dataflow stalls can arise from two primary situations:
(1) when the production rate exceeds the consumption rate, potentially filling the FIFO and causing stalling,
(2) or when the production rate is slower, leading to starvation in subsequent stages. Therefore, determining appropriate FIFO sizes is crucial for high performance.

We formulate the problem as follows.
Suppose the source stage can generate $C_{src}$ outputs per $II_{src}$ cycles, where $II_{src}$ is the initiation interval of the previous stage.
The destination stage demands $C_{dst}$ inputs per $II_{dst}$ cycles for computation.
The communication volume between the two stages is denoted as $V$.
We have functions, $f_{prod}$ (production rate) and $f_{con}$ (consumption rate), that track the number of samples generated and consumed at any given time $t$.
If there is no data in the FIFO, the consequential stage cannot perform computation since it does not receive any data.
Therefore, the consumption rate is always smaller or equal to the production rate.

\begin{minipage}{0.5\linewidth}
\begin{align}
\label{eq:prod}
f_{prod}(t)&=
\begin{cases}
C_{src}\lfloor t/II_{src}\rfloor & t\leq V/C_{src}II_{src}\\
V&t>V/C_{src}II_{src}
\end{cases}\\
\label{eq:con}
f_{con}(t)&=\max\left(C_{dst}\lfloor t/II_{dst}\rfloor,f_{prod}(t)\right)
\end{align}
\end{minipage}
\begin{minipage}{0.47\linewidth}
\includegraphics[width=\linewidth]{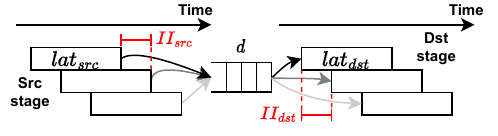}
\end{minipage}

Therefore, the FIFO depth between the source and destination stages can be calculated as:
\begin{equation}
\label{equ:fifo_depth}
d=\max_t\left(f_{prod}(t)-f_{con}(t)+1\right),\quad t\in\left[0,\argmin_{t'}\mathrm{abs}\left(V-\sum_{t'}f_{con}(t)\right)\right]\,,
\end{equation}
where the maximum $t$ represents the time required to receive all inputs.
$II_{src}$ and $II_{dst}$ can be obtained by running a high-level synthesis process.

This method is effective for optimizing a single connection between two stages.
However, in cases involving multiple stages, the production rate of the previous stages may influence all subsequent stages.
Based on Equation~\ref{eq:con}, let $f_{con}(t)=f_{prod}(t)$, we can obtain the new $II_{dst}'=II_{src}C_{dst}/C_{src}$.
This information is then propagated through the dataflow graph from the top down.
In cases where multiple producers feed into a single consumer, the resulting $II$ is the maximum of them, as the slowest stage dictates the overall pace.
Thus, we can determine the proper FIFO sizes for the entire dataflow graph.
Notably, this optimization only eliminates the first type of dataflow stalling but does not address potential design issues inside a kernel that might lead to a second type of stalling.


\section{Implementation}
\label{sec:impl}
\Name is implemented with 9K lines of Python code for the frontend ADL and 10K lines of C++ code for the \MLIR dialect and backend code generation.
In this section, we provide the implementation details of the frontend, type system, and codegen.

\paragraph{Frontend.}
\Name supports both imperative and declarative programming.
Unlike several schedule languages~\cite{chen2018tvm,lai2019heterocl} that rely on tracing-based techniques to generate AST and the corresponding IR, \Name utilizes Python's AST that provides the ability to handle control flow effectively.
Therefore, \Name can seamlessly accommodate the latest Python language features and remain compatible with the vast Python library ecosystem.
Consequently, \Name kernels can be executed with the native Python runtime to verify functional correctness, with minimal effort required to migrate a Python program into the \Name representation.

To accommodate larger designs, such as neural networks, \Name offers direct support for importing vision models from TorchVision~\cite{torchvision2016} and language models from HuggingFace~\cite{wolf2019huggingface} to achieve maximum flexibility.
\Name here serves as an intermediate language, showing its generality for hardware accelerator design.
We provide a backend for TorchDynamo~\cite{torchdynamo} in PyTorch 2.0~\cite{pytorch20}, so users can call \code{torch.compile(model,"allo")} to invoke the \Name compiler.
We employ \texttt{torch.fx}~\cite{reed2022fx} as the high-level IR and translate each PyTorch operator into a library function call within \Name.
PyTorch-level optimizations (e.g., operator fusion) are orthogonal to \Name's optimization.
As long as the model can be represented in \texttt{torch.fx}, \Name can take in and perform hardware-specific customizations.
Compared to writing \Name kernels in Python, this approach eliminates the need for users to rewrite the model and construct the schedule themselves.
The PyTorch frontend further simplifies programming by allowing users to directly import a model and utilize the high-performance \Name schedule out-of-the-box.
As our IR is constructed on top of \MLIR, we also plan to support other frontends within the \MLIR ecosystem~\cite{moses2021polygeist,iree,vasilache2022composablemlir} in the future.

\paragraph{Type System.}
\Name is equipped with a type inference engine designed to manage both built-in and custom data types.
The \Name type system differs from the Python native one, as it includes arbitrary bitwidth integers, fixed-point types, and additional shape information in the type hints.
\Name's type system consistently prevents overflow for any-bitwidth integers and fixed-point numbers, promoting data types with larger bitwidths when necessary.
Based on the predefined typing rules, the type inference engine starts from the annotations at the top-level function and tries to infer the data types of each inner variable.
In cases where the inferred data type deviates from the user's annotations, the engine attempts automatic type conversion if it is deemed feasible.
Furthermore, \Name incorporates shape information within the type declaration, facilitating shape inference, and thus supporting array slicing and automatic broadcasting.

\paragraph{Code Generation.}
After customizing the program, users can call \code{s.build(<target>)} to generate a valid program for CPU simulation or FPGA bitstream.
For the CPU backend, \Name lowers custom operations and data types to LLVM IR and uses the Just-in-Time execution engine~\cite{chris2021mlir} to run the program.
For the FPGA backend, \Name generates HLS C++ code for AMD Vivado/Vitis HLS~\cite{vitis_hls_2022}.
Since these tools accept programs written in C/C++, \Name directly generates code from the \texttt{affine} and \texttt{memref} dialects, bypassing the need for further lowering to lower-level dialects.
Annotated attributes such as pipelining and unrolling are converted into HLS pragmas during code generation.

\section{Experiments}
\label{sec:exp}
In this section, we first present our experiment settings and evaluate \Name against several baselines on a comprehensive benchmark and large neural network models.

\subsection{Experiment Settings}
For single-kernel evaluation, we compare \Name with ScaleHLS~\cite{yehpca2022scalehls}, PyLog~\cite{huang2021pylog}, HeteroCL~\cite{lai2019heterocl}, Merlin~\cite{merlin}, and Dahlia~\cite{rachit2020dahlia}, all of which generate HLS C++ code as output.
They represent the state-of-the-art ADLs and compilers that are publicly available.
We evaluate them on PolyBench~\cite{pouchet2012polybench}, a C-based benchmark suite consisting of commonly used kernels in scientific computing.
All experiments use the standard medium problem size and \texttt{float32} data types.

For multi-kernel evaluation, we evaluate three different convolutional neural networks (CNNs): ResNet-18~\cite{he2016resnet}, VGG~\cite{simonyan2014vgg}, and MobileNet~\cite{howard2017mobilenets}.
These models are implemented in PyTorch~\cite{paszke2019pytorch} and imported from the TorchVision~\cite{torchvision2016} library.
We run model inference and compare the results with ScaleHLS, which is the only frontend providing direct model import from PyTorch.
Other ADLs listed in Table~\ref{tab:comparison} do not provide Python bindings~\cite{koeplinger2018spatial,durst2020aetherling,thomas2020fleet} and do not generate HLS C++ code for backend synthesis.
Thus, it is challenging to reimplement these designs in their input languages, especially for large deep neural networks, making a fair comparison difficult.

To demonstrate the practical feasibility of \Name in generating large-scale designs running on real hardware, we implement an accelerator for the GPT2~\cite{radford2019gpt2} model, the only open-sourced model in the GPT family.
GPT2 is a Transformer-based, decoder-only architecture widely used in text generation tasks, with 355M parameters, 24 hidden layers, 16 heads in the attention module, and a hidden size of 1024.
We quantize the model into 4-bit weight and 8-bit activation (W4A8) for efficient deployment~\cite{frantar2022gptq,xiao2023smoothquant,zhao2023atom}, and verify the results against the quantized model in PyTorch to maintain accuracy.
We run backend synthesis for the design generated by \Name and deploy the bitstream on an FPGA.
For accelerators of such scale, all of the baseline ADLs fail to generate valid designs that satisfy the resource constraint.
Even when attempts are made to reduce the size of these designs, they still run into errors in the routing stage due to excessive memory access, leading to lengthy on-board wires.
Consequently, we directly compare \Name with DFX~\cite{hong2022dfx}, a state-of-the-art Transformer accelerator written in SystemVerilog.
We further compare the accelerator with two GPU devices, the NVIDIA GeForce GTX 1080Ti GPU, a widely-used commercial GPU, and the NVIDIA Tesla A100 GPU, a high-end GPU commonly employed for large-scale model training and inference.
Note that GPU requires ``fake'' quantization, so the actual low-bit performance is lower than the optimized fp16 performance, especially for models with less than one billion parameters~\cite{dettmers2022llmint8}.
Therefore, we report the best fp16 performance for GPUs in our experiments.

All the experiments target the AMD Alveo U280 FPGA using Vitis HLS v2022.1~\cite{vitis_hls_2022}.
The U280 FPGA has 4032 BRAM 18K blocks, 9024 DSP slices, 2.6M flip-flops, 1.3M LUTs, and 960 URAM blocks.
\Name generates the host program for Vitis in OpenCL.
AMD Xilinx RunTime (XRT) and Xilinx Board Utility (xbutil) are used for hardware execution and power measurements.
We set the frequency as \SI{300}{MHz} for high-level synthesis and \SI{250}{MHz} for on-board evaluation, since the large designs may have routing issues when we further increase the frequency.

\subsection{Single-Kernel Evaluation}
\label{sub:evaluation}
\begin{figure}[t]
    \centering
    \includegraphics[width=\textwidth]{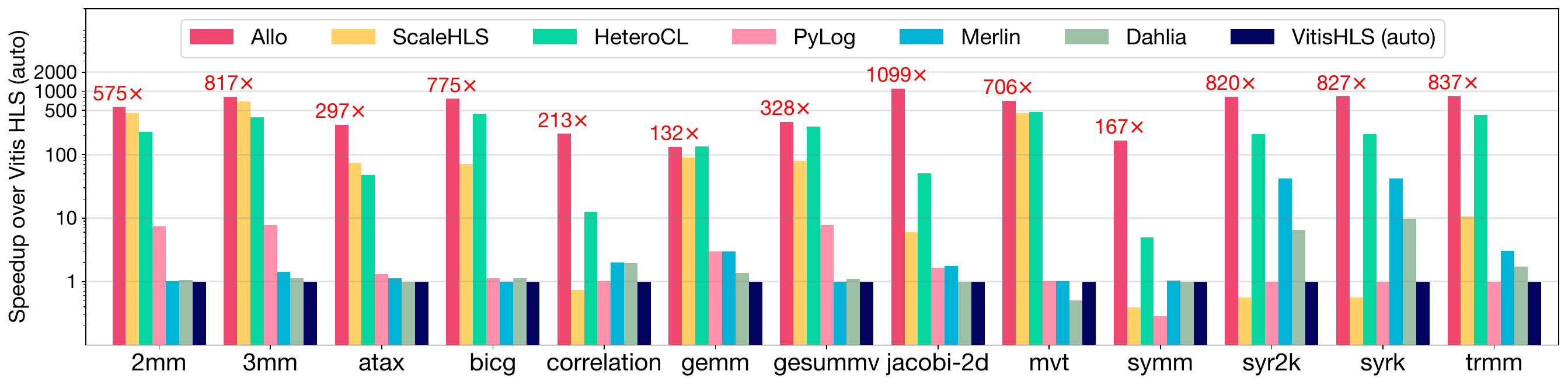}
    \vspace*{-20pt}
    \Description{PolyBench latency speedup over Vitis HLS baseline.}
    \caption{\textbf{PolyBench latency speedup over Vitis HLS baseline.}}
    \label{fig:polybench}
\end{figure}

Fig.~\ref{fig:polybench} shows the latency speedup against the Vitis HLS baseline for multiple ADLs. For Vitis HLS,  we do not apply any pragma or hardware customizations other than the default automatic loop pipelining. Merlin automatically inserts pragmas. ScaleHLS searches for the best optimizations from the Pareto-optimal frontier with a design space exploration (DSE) engine. We manually design the best optimization schemes with the available customization primitives for \Name, HeteroCL, PyLog, and Dahlia. Although PyLog provides automatic pragma insertion, we find manual customizations achieve better performance.
HeteroCL features kernel-level decoupled customizations with a lack of support for inter-kernel optimizations.
PyLog has a limited set of customization primitives compared to HeteroCL. For example, PyLog supports compute customizations such as loop unroll, pipeline, reorder, and tiling, but cannot build a custom memory hierarchy. Dahlia focuses on the predictability of HLS results instead of performance optimizations. For instance, Dahlia guarantees consistent memory banking and loop unrolling factors but lacks important support for loop pipelining.
\Name delivers up to 1099$\times$ latency speedup over Vitis HLS, 1478$\times$ over ScaleHLS, 34$\times$ over HeteroCL, 837$\times$ over PyLog, 775$\times$ over Merlin, and 1405$\times$ over Dahlia across different cases.

\begin{table}[!htbp]
\centering
\caption{\textbf{Results comparison between \Name and ScaleHLS} --- II denotes the loop pipeline initiation interval. We report the worst II for designs with multiple pipelined loops. We compare the clock frequency after placement and routing (PnR). The compilation time only includes the time to generate the HLS code.}
\label{tab:polybench}
\begin{adjustbox}{width=\textwidth}
\begin{tabular}{l|llllll|lllll}
\Xhline{1pt}
\multicolumn{1}{c|}{\multirow{3}{*}{Benchmark}} & \multicolumn{6}{c|}{\thead{\Name}}                                                                       & \multicolumn{5}{c}{\thead{ScaleHLS}}    \\ \cline{2-12} 
\multicolumn{1}{c|}{}  & Latency                                            & \multirow{2}{*}{II} & DSP                         & PnR             & Lines of    & Compile    & Latency    & \multirow{2}{*}{II} & DSP   & PnR           & Compile \\ 
\multicolumn{1}{c|}{}  & (cycles)                                           &                     & Usage                       & Freq. (MHz)     & \Name Custm.    & Time (s)   & (cycles)   &                     & Usage & Freq. (MHz)   & Time (s)\\ \Xhline{1pt}
atax                   & 4.9K (\textcolor{blue}{$\downarrow3.9\times$})     & 1                   & 403 (\textcolor{blue}{$\uparrow2.9\times$})   & 411             & 9                   & 1.0        & 19.4K      & 4                   & 141   & 329           & 36.1    \\
correlation            & 498.7K (\textcolor{blue}{$\downarrow290.5\times$}) & 1                   & 4168 (\textcolor{blue}{$\uparrow38.2\times$}) & 362             & 19                  & 0.8        & 144.9M     & 667                 & 109   & 305           & 638.8   \\
jacobi-2d              & 58.8K (\textcolor{blue}{$\downarrow183.1\times$})  & 1                   & 3968 (\textcolor{blue}{$\uparrow72.1\times$}) & 411             & 17                  & 0.9        & 10.8M      & 28                  & 55    & 308           & 47.9    \\
symm                   & 405.7K (\textcolor{blue}{$\downarrow427.4\times$}) & 1                   & 1208 (\textcolor{blue}{$\uparrow201.3\times$})& 402             & 15                  & 1.0        & 182.4M     & 13                  & 6     & 397           & 3.5     \\
trmm                   & 492.6K (\textcolor{blue}{$\downarrow78.0\times$})  & 1                   & 101  (\textcolor{blue}{$\uparrow14.4\times$}) & 414             & 12                & 0.8        & 38.4M      & 4                   & 7     & 382           & 1.4     \\ \Xhline{1pt}
\end{tabular}
\end{adjustbox}
\end{table}

We select five designs from the PolyBench suite where \Name outperforms ScaleHLS by a significant margin.
From Table~\ref{tab:polybench}, we see although ScaleHLS uses an automatic design-space exploration (DSE) engine to search for the best optimizations, the pipeline II of the DSE results is still high. A high pipeline II hurts the design performance in two ways: (1) the overall latency increases, and (2) HLS generates large multiplexers due to DSP reuse. These large multiplexers become critical paths and degrade the clock frequency. The frequency deterioration is especially evident for \texttt{atax}. ScaleHLS cannot fully pipeline the designs because of loop-carried dependency, excessive memory access, and complex nested loop structures of a monolithic kernel. For instance, \texttt{atax} has two matrix-vector products with loop-carried dependency. We implement the row-wise product discussed in \S\ref{sub:motivation_single} with \Name to break the loop-carried dependency and pipeline the inner loop to II=1. \texttt{jacobi-2d} is a typical stencil kernel with a sliding window data access pattern. ScaleHLS cannot fully pipeline such cases because it does not support data reuse. \Name builds two-level reuse buffers which drastically reduces off-chip memory access and fully pipelines the design. For \texttt{correlation}, \texttt{symm}, and \texttt{trmm}, \Name composes smaller kernels each customized with the aforementioned optimizations and forms a dataflow-pipelined design. The scale factor of DSP usage and the latency speedup are marked blue in Table~\ref{tab:polybench}. The fully pipelined designs customized with Allo deliver higher performance per DSP while attaining higher clock frequencies. In addition, \Name improves accelerator design productivity with fewer lines of code and short compilation time. The decoupled customizations and the kernel composition are expressed succinctly in less than 20 lines of code with \Name primitives.
We present full examples of customized \texttt{gemm} and \texttt{jacobi-2d} designs in Supplementary Material \textcolor{ACMPurple}{C}.


\subsection{Multi-Kernel Evaluation}
\label{sub:on_board}
\begin{figure}[t]
    \centering
    \includegraphics[width=0.75\linewidth]{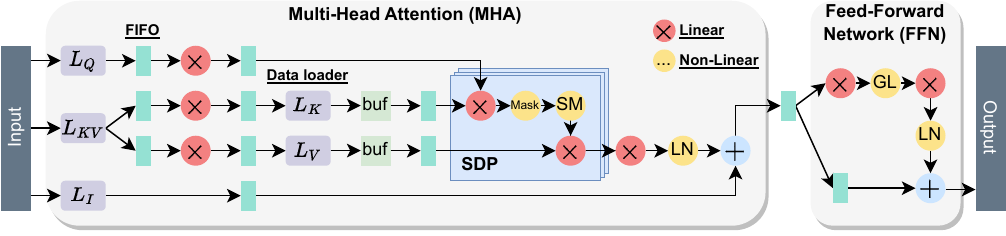}
    \caption{\textbf{A spatial architecture for the GPT2 model generated by \Name.}}
    \Description{A spatial architecture for the GPT2 model generated by \Name.}
    \label{fig:llm-arch}
\end{figure}

\subsubsection{Convolutional Neural Networks.}
Table~\ref{tab:nn} shows the results for CNNs.
\Name can achieve up to 12.7$\times$ speedup compared to ScaleHLS while remaining smaller resource usage in most cases.
This is because ScaleHLS does not perform kernel fusion and lacks support for reuse buffers.
\Name incorporates these optimizations to enhance data locality significantly.
Additionally, ScaleHLS falls short in fully exploiting the inter-kernel dataflow of designs, missing opportunities for optimization with FIFOs.
In contrast, \Name excels in efficiently leveraging dataflow characteristics, resulting in improved performance and resource utilization.
Our evaluations showcase the proficiency of \Name in composing diverse kernels to form a hierarchical structure and optimizing their orchestration, presenting a great advantage over the challenges encountered by ScaleHLS.
\begin{table}[!htbp]
\centering
\caption{\textbf{\Name and ScaleHLS per-sample latency and FPGA resource usage on CNN models} --- Vitis HLS automatically implements the Allo buffers using LUTs thus leading to zero BRAM usage.}
\label{tab:nn}
\resizebox{\linewidth}{!}{
\begin{tabular}{l|llllll|lllll}
\Xhline{1pt}
\multirow{3}{*}{Benchmark} & \multicolumn{6}{c|}{\thead{\Name}}     & \multicolumn{5}{c}{\thead{ScaleHLS}}    \\ \cline{2-12} 
\multicolumn{1}{c|}{}      & Latency (cycles)     & FF        & LUT     & BRAM     & DSP      & Speedup                          & Latency (cycles)     & FF      & LUT      & BRAM & DSP  \\ \Xhline{1pt}
VGG16                      & 3.85M                 & 36K     & 98K     & 0      & 440      & \textcolor{blue}{7.4$\times$}     & 28.31M                & 100K    & 714K     & 3936 & 882  \\
MobileNet                  & 0.26M                &  57K      & 128K    & 0      & 1942     & \textcolor{blue}{8.3$\times$}     & 2.17M                 & 93K    & 518K      & 6796 & 1778 \\
ResNet18                   & 8.29M                 & 51K    & 124K     & 0      & 652      & \textcolor{blue}{12.7$\times$}    & 104.88M               & 144K    & 992K     & 8416 & 1330  \\ \Xhline{1pt}
\end{tabular}
}
\vspace{-.1in}
\end{table}

\subsubsection{Large Language Models.}
The accelerator architecture for the GPT2 model is depicted in Fig.~\ref{fig:llm-arch}.
A typical Transformer block consists of an MHA module that leverages scaled dot-product (SDP) to calculate the attention score and an FFN that cascades two linear layers.
LLMs like GPT2 are inherently more complex than CNNs, both in terms of model sizes and the intricate connections within the MHA module that require splitting the heads and merging the results at the end.
Existing ADLs and compilers often lack explicit memory and dataflow management~\cite{lai2019heterocl,yehpca2022scalehls,huang2021pylog}.
Therefore, the designs they generate often demand significantly more resources – exceeding twice the on-chip resources available.
In contrast, \Name enables the design and optimization of each submodule individually, providing control over memory hierarchy and data orchestration strategies.
After verifying their correctness, these submodules can be composed bottom-up using the \code{.compose()} primitive, and later connected with \code{.relay()} primitive to form a complete design.

\begin{figure}[t]
\begin{minipage}{0.67\linewidth}
\centering
\includegraphics[width=\linewidth]{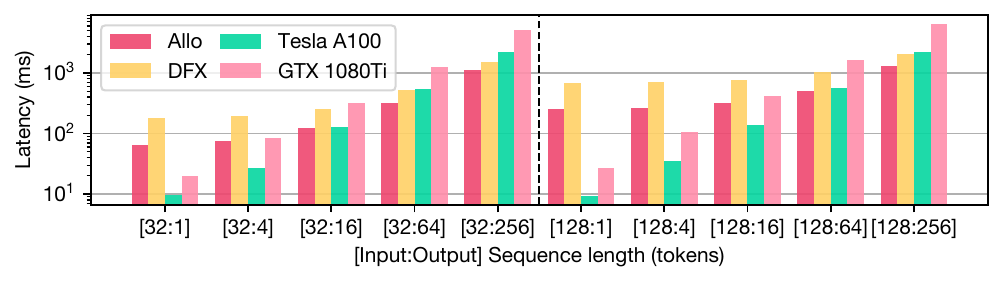}
\end{minipage}
\begin{minipage}{0.3\linewidth}
\resizebox{\linewidth}{!}{
\begin{tabular}{ccc}\Xhline{1pt}
 & \textbf{Allo} & \textbf{DFX}\\\Xhline{1pt}
Device & U280 & U280\\
Freq. & 250MHz & 200MHz\\
Quant. & W4A8 & fp16\\
BRAM & 384 (19.0\%) & 1192 (59.1\%)\\
DSP & 1780 (19.73\%) & 3533 (39.2\%)\\
FF & 652K (25.0\%) & 1107K (42.5\%)\\
LUT & 508K (39.0\%) & 520K (39.9\%)\\\Xhline{1pt}
\end{tabular}
}
\end{minipage}
\vspace{-5pt}
\caption{Left: \textbf{End-to-end latency on GPT2.} Right: \textbf{Resource utilization of \Name and DFX.}}
\Description{Left: \textbf{End-to-end latency on GPT2.} Right: \textbf{Resource utilization of \Name and DFX.}}
\label{fig:latency}
\vspace{-8pt}
\end{figure}

In this experiment, we consider the generative inference scenario for LLMs in a single-batch low-latency setting~\cite{chen2024llmfpga}.
We adjust the input and output sequence lengths and measure the end-to-end latency from kernel launch, including the CPU-GPU/FPGA communication time.
As depicted in Fig.~\ref{fig:latency}, \Name consistently outperforms the state-of-the-art accelerator DFX, achieving up to 2.80$\times$ speedup in terms of latency.
This is primarily attributed to the highly customized high-performance systolic array kernels and the efficient composition of multiple kernels within \Name.
\Name also utilizes fewer resources than DFX from the right side of Fig.~\ref{fig:latency}.
This is because the spatial architecture designed in \Name maximally reduces on-chip intermediate buffers.
In contrast, DFX employs an overlay design that reuses hardware units for various operators, thereby increasing resource utilization.
Thus, we can achieve a higher frequency but use fewer resources than DFX.
Notably, the GPT2 model is directly imported using the PyTorch frontend (\S\ref{sec:impl}), which does not require users to rewrite any code for the model.
The accelerator fully leverages the \Name customization primitives with less than 50 lines of customization code to optimize this intricate design, which is much more productive than implementing it in a hardware description language (HDL).

Furthermore, we extend our performance evaluation to compare with two GPU devices.
The FPGA-based design exhibits a notable performance gap compared to GPUs when the output sequence length is small.
This discrepancy is primarily due to the extensive compute requirements during the initial stage of generative inference, known as the prefill stage, which aligns more effectively with throughput-oriented devices such as GPUs~\cite{pope2022googleinf,chen2024llmfpga}.
However, as the output sequence length increases, the FPGA accelerator generated by \Name surpasses GPU performance.
In particular, we achieve a 5.05$\times$ speedup compared to the 1080Ti GPU.
Even in comparison to a high-end A100 GPU, we still attain a 1.70$\times$ speedup for longer output sequences.
Additionally, the FPGA execution requires only 30 watts of measured power, whereas the A100 counterpart demands 96 watts, which means the \Name accelerator is 5.44$\times$ more energy-efficient than the A100 GPU.


\section{Related Work}
\label{sec:related}
\paragraph{Schedule Languages.}
Halide~\cite{jrk2013halide} first introduces the concept of algorithm and schedule decoupling in the domain of image processing.
TVM~\cite{chen2018tvm} extends this idea to deep learning and supports end-to-end optimization workflow mapping neural network models to different hardware devices.
There are also other DSLs that leverage schedule languages to generate high-performance code in their specific domains~\cite{baghdadi2019tiramisu,zhang2018graphit,chen2024slapo,hagedorn2020fireiron,fredrik2017taco,vasilache2018tensorcomp,lai2019heterocl,krill2021chen}.
\Name also adopts this decoupling idea but further enhances the composability of customizations and schedules.
It leverages a hierarchical dataflow graph to compose smaller designs into larger ones, achieving high performance on large designs.

Recent developments explore program rewriting techniques as an alternative to the traditional schedule tree, making it possible to handle more imperative programs.
For instance, TensorIR~\cite{feng2022tir} extends TVM to support a more flexible syntax for describing computations, enabling better tensorization for TensorCore on GPUs.
Exo~\cite{yuka2022exo} formalizes program rewrite rules using an effect system to guarantee the correctness of transformations.
The \texttt{xform} dialect~\cite{xform} in \MLIR also supports rewrites to the programs.
While these efforts mainly focus on kernel-level optimizations, they lack the capability to effectively compose optimizations across multiple kernels, limiting their scalability to larger and more complex designs.

\paragraph{Accelerator Design Languages (ADLs) and Compilers.}
Numerous domain-specific languages (DSLs) have emerged to facilitate hardware designs for different applications~\cite{duarte2018particle,milder2012spiral,steward2018ripl,hegarty2014darkroom,reiche2017hipacc,hegarty2016rigel,li2020heterohalide,wang2017p4fpga,hardik2016dnnweaver,debjit2022dac}, providing highly optimized operators tailored for specific domains.
Subsequently, various ADLs have been introduced to address more general-purpose accelerator design~\cite{koeplinger2016dhdl,koeplinger2018spatial,tal2019dace,huang2021pylog,durst2020aetherling,thomas2020fleet}.
However, their algorithm and customizations are entangled together, leading to reduced productivity and limited exploration of different customization combinations.
HeteroCL~\cite{lai2019heterocl} decouples hardware customizations from the algorithm but primarily focuses on single-kernel designs.
Several ADLs also emphasize dataflow optimizations~\cite{durst2020aetherling,thomas2020fleet,xiang2022heteroflow,koeplinger2018spatial,tal2019dace}, yet they struggle to preserve the hierarchical structure of dataflow and cannot efficiently compose small kernels into larger designs.
Consequently, they encounter challenges when scaling to accommodate large and complex models.

Recent efforts harness the \MLIR toolchain to generate C/C++ HLS code~\cite{yehpca2022scalehls,zhao2022polsca,bohm2022soda}.
However, the existing compiler passes and design space exploration (DSE) engines often fall short in producing high-performance accelerators, as showcased in Section \ref{sub:evaluation}.
This is primarily because \MLIR lacks inherent support for crucial components like quantized data types, memory, and dataflow customizations.
Lastly, low-level hardware design languages (HDLs) such as Calyx~\cite{rachit2021calyx} are designed to facilitate the process of backend synthesis.
Filament~\cite{nigam2023filament} is also a low-level HDL that leverages timeline types to reason about timing safety.
These efforts are orthogonal to ours, and we plan to support the CIRCT~\cite{CIRCT} project as a backend in the future.

\section{Conclusion and Future Work}
\label{sec:conclusion}
In this paper, we propose \Name, a composable programming model for accelerator design.
\Name proposes progressive hardware customizations allowing users to apply provable program transformations step by step, and further introduces composable schedules to combine small kernels into large designs.
Nonetheless, there are several unexplored directions in our ongoing work.

For optimizations within a kernel, we plan to design an autoscheduler that can reduce the programming efforts required from developers.
For composing multiple kernels, the techniques proposed in \S\ref{sub:fifo_sizing} only address FIFO sizing but do not determine where to establish these connections automatically.
Some kernels may have dependency relations that prevent direct connection with FIFOs and might require additional buffers to ensure sequential memory access.
We plan to develop automatic bufferization techniques to create buffers between stages and guarantee correctness.

Practical hardware design entails more than just customizing and transforming code; it also involves connecting components and guiding them through the entire backend synthesis process to generate a bitstream.
Mapping dataflow regions onto multi-die FPGA boards can be challenging.
Several large designs we experimented with for LLMs failed to meet timing requirements, often due to issues during the routing stage.
In order to improve the frequency of the design, it is essential to explicitly bind dataflow regions to specific hardware regions and minimize cross-die communication.
Although efforts such as AutoBridge~\cite{guo2021autobridge} aim to decompose designs into smaller parts and assemble them, they cannot accommodate complex hierarchical dataflow or create double buffers.
We plan to create a build system that can compile the entire design in parallel and efficiently link the components together, similar to how software linkers work.
This approach will help optimize the hardware design process further and enhance performance.

\newpage
\begin{acks}
This work was supported in part by ACE, one of the seven centers in JUMP 2.0, a Semiconductor Research Corporation (SRC) program sponsored by DARPA and NSF Awards \#2019306 and \#2118709, and an Intel ISRA award. 
We gratefully acknowledge the anonymous reviewers, Prof. Adrian Sampson, Prof. Louis-No\"el Pouchet, Rachit Nigram, and Jialu Bao for their valuable feedback on the initial draft of this work. 
We thank Jiahao Zhang for providing a reference LLM accelerator implementation.
We also thank Jin Yang, Jeremy Casas, and Zhenkun Yang for their insightful feedback on the initial  version of the \Name framework.
\end{acks}

\section*{Artifact}
The \Name code is open-source and available at the \href{https://github.com/cornell-zhang/allo}{allo repository} on GitHub. Detailed instructions for reproducibility and reusability are provided in an archived version on Zenodo~\cite{zenodo_allo} and at the \href{https://github.com/cornell-zhang/allo-pldi24-artifact}{allo-pldi24-artifact} repository on GitHub.

\bibliographystyle{ACM-Reference-Format}
\bibliography{pldi24}

\newpage
\pagestyle{empty}
\appendix
\section{Formal Data Type Annotation}
\label{appendix:type}
Fig.~\ref{fig:types} shows the grammar of type annotation in \Name.
\Name supports different Python builtin types, any-bitwidth integer types, and fixed-point data types.

\begin{figure}[!htbp]
\hfill
\begin{minipage}{0.36\linewidth}
\begin{center}
\begin{minted}[xleftmargin=\parindent,
               fontsize=\scriptsize]{python}
T ::= eleTy | eleTy [ shape ]
shape ::= dim | dim , shape
dim   ::= [0-9]+ | constvar | expr
expr  ::= dim binop dim
binop ::= + | - | * | /
eleTy ::= intTy | floatTy | fixedTy
\end{minted}
\end{center}
\end{minipage}
\vline\hfill
\begin{minipage}{0.55\linewidth}
\begin{center}
\begin{minted}[xleftmargin=\parindent,
               fontsize=\scriptsize]{python}
intTy ::= int[0-9]+ | uint[0-9]+ | index
floatTy ::= float16 | float32 | float64
fixedTy ::= fixed(width, frac) | ufixed(width, frac)
constvar ::= [a-zA-Z_][a-zA-Z0-9_]*
width ::= [0-9]+
frac  ::= [0-9]+
\end{minted}
\end{center}
\end{minipage}
\caption{Context-free grammar of basic data types of \Name.}
\Description{Context-free grammar of basic data types of \Name.}
\label{fig:types}
\end{figure}

\section{Memory Layout Propagation Termination Theorem}
\label{appendix:layout}
\begin{theorem}
Algorithm~\ref{alg:layout} can terminate in $\mathcal{O}(M)$ steps.
\end{theorem}
\begin{proof}
To apply the Fixed-Point Theorem, we need to prove (1) the lattice $\mathcal{L}$ of partition types for all the nodes is complete, and (2) The transfer function $\mathcal{F}:\mathcal{L}\to\mathcal{L}$, which represents the transition from one iteration to the next, is monotonic.

First, we decompose the partition type $t_i$ into $(t_{i0},\ldots,t_{iD_i})$, where $D_i$ is the number of dimensions for the $i$-th dataflow variable.
Thus, the partition types for all the dataflow nodes can be denoted as $V^{(k)}=(t_{00}^{(k)},\ldots,t_{0D_0}^{(k)},t_{10}^{(k)},\ldots,t_{MD_M}^{(k)})$ with the superscript indicating the $k$-th iteration.
Since each element of $V^{(k)}_i$ forms a complete lattice $\mathcal{L}_i$ based on the definition of Fig.~\ref{fig:lattice}, the product lattice $\mathcal{L}^{MD_M}=\mathcal{L}_0\times \cdots\times \mathcal{L}_{MD_M}$ is also finite and complete, which finishes the first part of the proof.

Based on Line 4 of Algorithm~\ref{alg:layout}, the meet operator $\sqcap$ is the only operator used in each operation.
Thus, the transfer function $\mathcal{F}:\mathcal{L}^{MD_M}\to\mathcal{L}^{MD_M}$ is always monotonic.
This ensures we meet the precondition for the Fixed-Point Theorem, guaranteeing that Algorithm~\ref{alg:layout} can terminate in a finite number of steps and reach a fixed point.
The fixed point can be obtained by plugging in $\top$ to the algorithm (no partition for all the variables) and iterating multiple times to obtain $\mathcal{F}^K(\top)$.

Lastly, the worst case of the algorithm is to go from the $\top$ of $\mathcal{L}^{MD_M}$ to the $\bot$, which requires at most $\mathrm{Depth}(\mathcal{L}^{MD_M})$ steps.
Since the number of dimensions for each dataflow node is typically a small constant ($\leq$ 4), and the depth of each lattice $\mathcal{L}_i$ is also a fixed constant not related to $M$, the final complexity of the algorithm is in linear time $\mathcal{O}(M)$.
\end{proof}

\section{Full \Name Program Examples}
\label{sec:examples}
The following code snippets show full examples of customized kernels with \Name.
The first example demonstrates how to transform an inner-product GEMM to a row-wise product GEMM in \Name, which reflects the example in Fig.~\ref{fig:single_hls}.
We can see \Name only needs three lines of customization code to achieve the transformation.

\begin{mdframed}[
  leftline=false,       
  rightline=false,      
  topline=false,        
  bottomline=false      
]
\begin{minted}[linenos,
               fontsize=\scriptsize]{python}
import allo
from allo.ir.types import float32
import numpy as np

M, N, K = 1024, 1024, 1024

# Algorithm specification
def gemm(A: float32[M, K], B: float32[K, N]) -> float32[M, N]:
    C: float32[M, N] = 0.0
    for i, j in allo.grid(M, N):
        for k in allo.reduction(K):
            C[i, j] += A[i, k] * B[k, j]
    return C

# Schedule construction
s = allo.customize(gemm)
s.reorder("k", "j")
s.buffer_at(gemm.C, axis="i")
s.pipeline("j")

# Build CPU executable
mod = s.build(target="llvm")
np_A = np.random.randint(0, 100, (32, 32)).astype(np.int32)
np_B = np.random.randint(0, 100, (32, 32)).astype(np.int32)
np_C = mod(np_A, np_B)
golden_C = np.matmul(np_A, np_B)
np.testing.assert_allclose(np_C, golden_C, rtol=1e-5, atol=1e-5)

# Genenerate FPGA bitstream
mod_hls = s.build(target="vitis", mode="hw", project="gemm.prj")
mod_hls()
\end{minted}
\end{mdframed}

\smallskip
The second example optimizes \texttt{jacobi-2d} to II=1 by building a two-level memory hierarchy with the \code{.reuse_at()} primitive.
It demonstrates how \Name composes customized kernels to form an optimized top-level design.

\begin{mdframed}[
  leftline=false,       
  rightline=false,      
  topline=false,        
  bottomline=false      
]
\begin{minted}[linenos,
               fontsize=\scriptsize]{python}
import allo
from allo.ir.types import float32

TSTEPS = 100
N = 250

# Algorithm specification
def compute_A(A0: float32[N, N], B0: float32[N, N]):
    for i0, j0 in allo.grid(N-2, N-2, name="A"):
        B0[i0, j0] = 0.2 * (
            A0[i0, j0 + 1] + A0[i0 + 1, j0] + A0[i0 + 1, j0 + 1] + A0[i0 + 1, j0 + 2] + A0[i0 + 2, j0 + 1]
        )

def compute_B(B1: float32[N, N], A1: float32[N, N]):
    for i1, j1 in allo.grid(N-2, N-2, name="B"):
        A1[i1 + 1, j1 + 1] = 0.2 * (
            B1[i1, j1 + 1] + B1[i1 + 1, j1] + B1[i1 + 1, j1 + 1] + B1[i1 + 1, j1 + 2] + B1[i1 + 2, j1 + 1]
        )

def kernel_jacobi_2d(A: float32[N, N], B: float32[N, N]):
    for m in range(TSTEPS):
        compute_A(A, B)
        compute_B(B, A)

# Customize compute_A
sch0 = allo.customize(compute_A)
lb0 = sch0.reuse_at(sch0.A0, "i0") # build line buffer
wb0 = sch0.reuse_at(lb0, "j0") # build window buffer
sch0.pipeline("i0")
sch0.partition(lb0, dim=0)
sch0.partition(wb0, dim=0)

# Customize compute_B
sch1 = allo.customize(compute_B)
lb1 = sch1.reuse_at(sch1.B1, "i1")
wb1 = sch1.reuse_at(lb1, "j1")
sch1.pipeline("i1")
sch1.partition(lb1, dim=0)
sch1.partition(wb1, dim=0)

# Compose customizations to the top-level schedule
sch = allo.customize(kernel_jacobi_2d)
sch.compose(sch0, sch1)    
sch.partition(sch.A, dim=2)
sch.partition(sch.B, dim=2)

# Generate FPGA bitstream
hls_mod = sch.build(target="vitis", mode="hw", project="jacobi2d.prj")
hls_mod()
\end{minted}
\end{mdframed}

\section{Example Generated HLS C++ Code}
\label{appendix:hls_cpp}
The following code shows an example of HLS C++ code generated by \Name for a systolic array.
This example demonstrates the complexity of HLS design, incorporating compute, memory, and communication customizations.
\begin{mdframed}[
  leftline=false,       
  rightline=false,      
  topline=false,        
  bottomline=false      
]
\begin{minted}[linenos,
               fontsize=\scriptsize]{cpp}
void kernel_0_0(
  hls::stream< int8_t > &v0, hls::stream< int8_t > &v1,
  hls::stream< int8_t > &v2, hls::stream< int8_t > &v3,
  int16_t v4[4][4],
  int v5, int v6
) {
  #pragma HLS stream variable=v0 depth=5
  #pragma HLS stream variable=v1 depth=5
  #pragma HLS stream variable=v2 depth=5
  #pragma HLS stream variable=v3 depth=5
  #pragma HLS array_partition variable=v4 complete dim=1
  #pragma HLS array_partition variable=v4 complete dim=2

  l_S_k_0_k: for (int k = 0; k < 4; k++) {
    int8_t v8 = v0.read();
    int8_t a;
    a = v8;
    int8_t v10 = v1.read();
    int8_t b;
    b = v10;
    int8_t v12 = a;
    int16_t v13 = v12;
    int8_t v14 = b;
    int16_t v15 = v14;
    int16_t v16 = v13 * v15;
    int16_t v17 = v4[v5][v6];
    int16_t v18 = v17 + v16;
    v4[v5][v6] = v18;
    int8_t v19 = a;
    v2.write(v19);
    int8_t v20 = b;
    v3.write(v20);
  }
}

void kernel_1_0(...)
// ... 16 PE kernels in total

void systolic_array(
  int8_t v336[4][4],
  int8_t v337[4][4],
  int16_t v338[4][4]
) {
  #pragma HLS dataflow
  #pragma HLS array_partition variable=v336 complete dim=1
  #pragma HLS array_partition variable=v337 complete dim=2
  #pragma HLS array_partition variable=v338 complete dim=1
  #pragma HLS array_partition variable=v338 complete dim=2
  hls::stream< int8_t > A_fifo[4][5];
  #pragma HLS stream variable=A_fifo depth=5
  hls::stream< int8_t > B_fifo[4][5];
  #pragma HLS stream variable=B_fifo depth=5
  l_data_load_k16: for (int k16 = 0; k16 < 4; k16++) {
    l_S_m_0_m: for (int m = 0; m < 4; m++) {
      int8_t v343 = v336[m][k16];
      A_fifo[m][0].write(v343);
    }
    l_S_n_1_n: for (int n = 0; n < 4; n++) {
      int8_t v345 = v337[k16][n];
      B_fifo[n][0].write(v345);
    }
  }
  hls::stream< int8_t > &v346 /* v346[4] */ = A_fifo[0][0];
  hls::stream< int8_t > &v347 /* v347[4] */ = B_fifo[0][0];
  hls::stream< int8_t > &v348 /* v348[4] */ = A_fifo[0][1];
  hls::stream< int8_t > &v349 /* v349[4] */ = B_fifo[0][1];
  kernel_0_0(v346, v347, v348, v349, v338, 0, 0);
  // invoke other 15 PE functions
  // ...
  int8_t A_drain[4];
  int8_t B_drain[4];
  l_data_drain_k17: for (int k17 = 0; k17 < 4; k17++) {
    l_S_m_4_m1: for (int m1 = 0; m1 < 4; m1++) {
      int8_t v414 = A_fifo[m1][4].read();
      A_drain[m1] = v414;
    }
    l_S_n_5_n1: for (int n1 = 0; n1 < 4; n1++) {
      int8_t v416 = B_fifo[n1][4].read();
      B_drain[n1] = v416;
    }
  }
}
\end{minted}
\end{mdframed}

\end{document}